\documentclass[14pt]{article}
\usepackage{latexsym}
\usepackage{amssymb}
\usepackage{amsmath}
\usepackage{graphicx}
\usepackage{epstopdf}
	\DeclareGraphicsExtensions{.eps}
\usepackage{multicol}
\usepackage{caption}
\usepackage[pagewise]{lineno}
\usepackage{subfig}
\usepackage[T1]{fontenc}
\usepackage[utf8]{inputenc}
\usepackage{authblk}
\usepackage{multirow}
\usepackage{multicol}
\usepackage{textcomp}
\usepackage[utf8]{inputenc}
\usepackage{array}
\usepackage{adjustbox}
\usepackage{booktabs}
\usepackage{xcoffins}

\newcommand{\V}{\mbox{V}}

\NewCoffin\tablecoffin
\NewDocumentCommand\Vcentre{m}
  {%
    \SetHorizontalCoffin\tablecoffin{#1}%
    \TypesetCoffin\tablecoffin[l,vc]%
  }
\begin{document}

\vspace*{1cm}
\begin{center}
\def\cl{\centerline}
{\large
{\bf \textbf{Probability density function of in-plane permeability of fibrous media: Constant Kozeny coefficient}}}
\vskip 1 truecm
\cl{\large{\bf M.~Bodaghi$^{a}$, S.~Yasaei Sekeh$^{b}$, N.~Correia$^{c,*}$}}
\vskip .5 truecm

\cl{$^{a}$ Engineering Design and Advanced Manufacturing, MIT Portugal Program,}
 \cl{Faculty of Engineering, University of Porto, Porto, Portugal }
\cl{$^{b}$ Department of Statistics, Federal\;University\;of\;S$\tilde{\rm a}$o\;Carlos, SP, Brazil}

\cl{$^{c}$ Institute of Mechanical Engineering and Industrial Managment,}
\cl{Faculty of Engineering, University of Porto, Porto, Portugal}

\author[1]{Author A\thanks{A.A@university.edu}}
\author[1]{Salimeh Yasaei Sekeh\thanks{sa$_{-}$yasaei@yahoo.com}}
\author[1]{Author C\thanks{C.C@university.edu}}
\author[2]{Author D\thanks{D.D@university.edu}}
\author[2]{Author E\thanks{E.E@university.edu}}
\affil[1]{Department of Computer Science, \LaTeX\ University}
\affil[2]{Department of Statistics, \LaTeX\ Federal University of S$\tilde{a}$o Carlos (UFSCar), BRAZIL}
\affil[3]{Department of Mechanical Engineering, \LaTeX\ University}
\bigskip

\end{center}
\bigskip



\footnotetext{$^*$ Corresponding author: E-mail: nuno.correia@inegi.up.pt, Tel: (+351-229578710)}

\begin{abstract}
Permeability of fibrous porous media at the micro/meso scale-level is subject to significant uncertainty due to the heterogeneity of the fibrous media. The local microscopic heterogeneity and spatial variability porosity, tortuosity and fibre diameter affect the experimental measurements of permeability at macroscopic level. This means that the selection of an appropriate probability density function (PDF) is of crucial importance, in the characterization of both  local variations at the microscale and the equivalent permeability at the experimental level (macroscale). This study addresses the issue of whether or not a normal distribution appropriately represents permeability variations. To do so, (i) the distribution of local fibre volume fraction for each tow is experimentaly determined by estimation of each pair of local areal density and thickness, (ii) the Kozeny-Carmen equation together with the change of variable technique are used to compute the PDF of permeability, (iii) using the local values of fibre volume fraction, the distribution of local average permeability is computed and subsequently the goodness of fit of the computed PDF is compared with the distribution of the permeability at microscale level. Finally variability of local permeability at the microscale level is determined.\\
 The first set of results reveals that (1) the relationship between the local areal density and local thickness in a woven carbon-epoxy composite is modelled by a bivariate normal distribution, (2) while fibre volume fraction follows a normal distribution, permeability follows a gamma distribution, (3) this work also shows that there is significant agreement between the analytical approach and the simulation results. The second set of results shows that the coefficient of variation of permeability is one order of magnitude larger than that of fibre volume fraction. Future work will consider other variables, such as   type of fabrics, the degree of fibre preform compaction to determine whether or not the bivariate normal model is applicable for a broad range of  fabrics.
\end{abstract}


\mbox{\quad}
\def\fB{\mathfrak B}\def\fM{\mathfrak M}\def\fX{\mathfrak X}
 \def\cB{\mathcal B}\def\cM{\mathcal M}\def\cX{\mathcal X}
\def\bu{\mathbf u}\def\bv{\mathbf v}\def\bx{\mathbf x}\def\by{\mathbf y}
\def\om{\omega} \def\Om{\Omega}
\def\bbP{\mathbb P} \def\hw{h^{\rm w}} \def\hwphi{{h^{\rm w}_\phi}}
\def\beq{\begin{eqnarray}} \def\eeq{\end{eqnarray}}
\def\beqq{\begin{eqnarray*}} \def\eeqq{\end{eqnarray*}}
\def\rd{{\rm d}} \def\Dwphi{{D^{\rm w}_\phi}}
\def\BX{\mathbf{X}}
\def\mwe{{D^{\rm w}_\phi}}
\def\DwPhi{{D^{\rm w}_\Phi}} \def\iw{i^{\rm w}_{\phi}}
\def\bE{\mathbb{E}}
\def\1{{\mathbf 1}} \def\fB{{\mathfrak B}}  \def\fM{{\mathfrak M}}
\def\diy{\displaystyle} \def\bbE{{\mathbb E}}
\def\bbR{\mathbb R}
\def\BC{{\mathbf C}} \def\bbR{{\mathbb R}} \def\bmu{{\mbox{\boldmath${\mu}$}}}
 \def\bPhi{{\mbox{\boldmath${\Phi}$}}}
 \def\bbZ{{\mathbb Z}} \def\diy{\displaystyle}

\section{Introduction}
Fibrous media display different degrees of meso-scale variability from ideal fibre paths. This can be due to the manufacturing of the reinforcement, handling and preparing the moulding step. Resin flow inside porous media is influenced by fibres spatial variability and heterogeneity, and neglecting this causes errors in process analysis and uncertainty in measurement. Thus a reliable model of fluid flow in heterogeneous media must include  multiscale phenomena and capture the multiscale nature of fluid transport behaviour, at microscale, mesoscale and macroscale. In the sense the dominant processes and governing equations may vary with scales. Therefore, extending from microscale level to a mesoscale one needs upscaling that allows the essence of physical processes at one level to be summarized at the larger level. However, a detailed understaning of  upscaling process from microscale to mesocale has not been completed. Mesoscopic and macroscopic properties of fibrous media such as porosity, fibre size distribution and permeability can be characterized through lab-scale methods while the microscale properties are uncaptured. The lack of ability for measurements on the microscale can lead to uncertainty in interpretations of the data captured at macroscale. A major challenge arising from this non homogeneity is how macroscale flow is influenced by the microscale structure (pore spaces), as well as by the physical properties of the resin.
Permeability is an important macroscale variable representing average of microscale properties of porous media. This average permeability is the fundamental property arising from Dracy's law (\ref{eq.Darcy}):
\beq \label{eq.Darcy} \overline{u}=\frac{K}{\mu}.\Delta p ,\;\;\; \hbox{ (1D version of the equation)} \eeq
which describes the relation between the volume averaged fluid velocity, \={u}, the pressure gradient, $\Delta$p, the fluid’s viscosity ,$\mu$, and the equivalent permeability tensor K.  Empirical equations, such as Kozeny$-$Carmen (\ref{eq.Kozeny}), have been developed to relate the meso-scale permeability with the microscale properties, including porosity:
\beq \label{eq.Kozeny} K=\frac{r_f^2}{k_c}\frac{(1-V_f)^3}{(V_f)^2}.\eeq
Here $k_c=C\tau^2$ and $\tau^2=\frac{L_e}{L}$,
 where $V_f$ is porosity, $ r_f$ is fibre radius, $ k_c$ is Kozeny constant, $\tau$ is tortuosity, $ L_e $ is the  length of streamlines, L is the length of sample and C is a proportionality constant \cite{CCP}.

\begin{figure}[h!]
\centerline{\includegraphics[scale=.8]{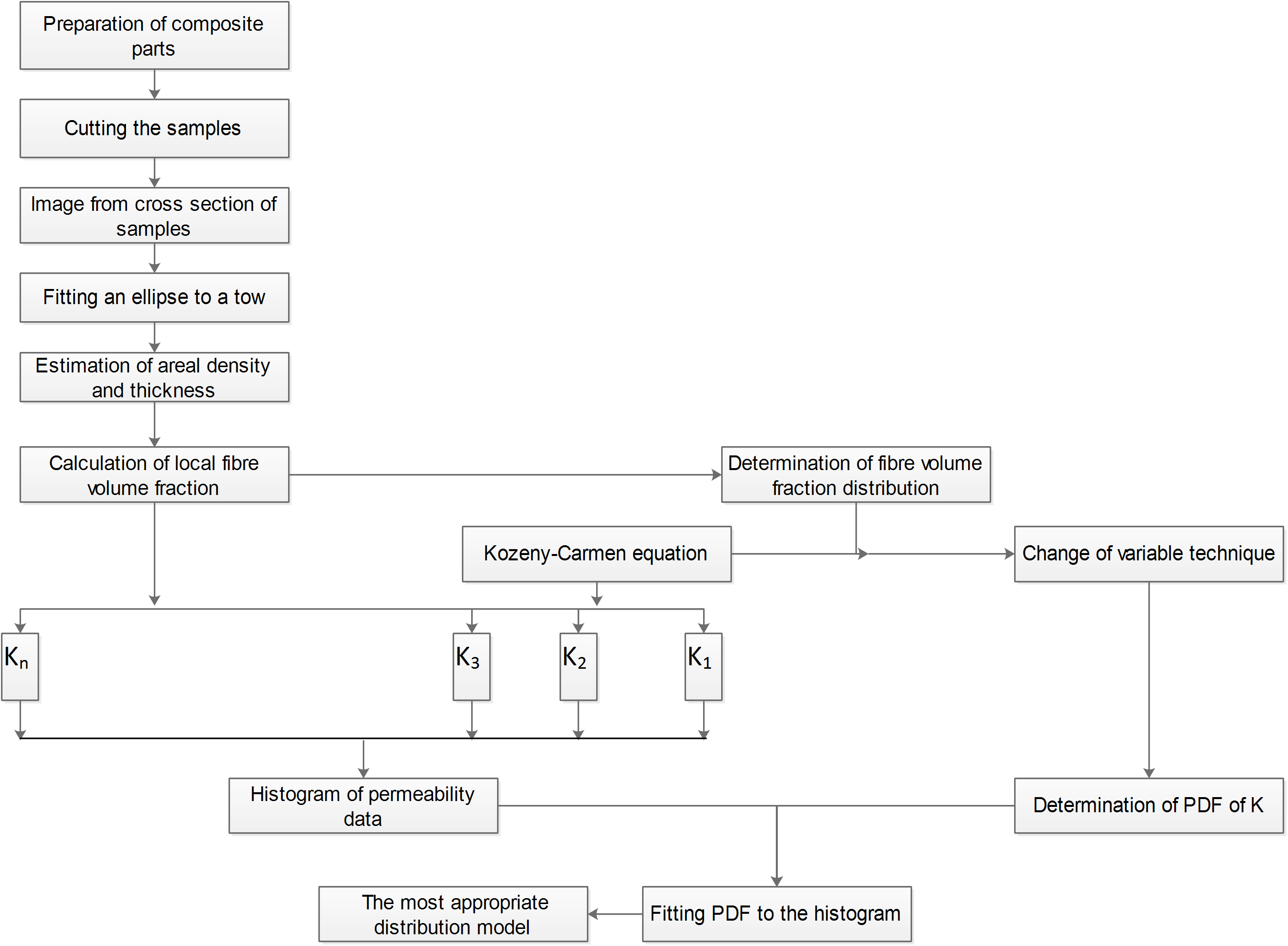}}
\caption{Flowchart for probability density function of permeability}
\label{fig:flowchart1}
\end{figure}

Because the nonhomogeneous nature of porous media originates in the randomness of fibre diameter distributions, porosities and pore structure, permeability is subjected to uncertainty \cite{M}. Causes and effects of this uncertainty have been reviewed \cite{U,B},  assuming a normal \cite{B,I,II,S,H,SS,SP,SPW} and a lognormal permeability probability density function \cite{SM}, previous studies have modeled the effect of this uncertainty on fluid flow in porous media by employing stochastic analysis. Considered as a vast area in stochastic processes, such analyzes require to identify the sources of uncertainties and to select probabilistic methods for uncertainty propagation up to different modeling levels.\\
 In simulation of mould filling process, such as Resin Transfer Moulding, which are described by Dracy’s law (\ref{eq.Darcy}), permeability directly affects filling time and flow pattern. An accurate probability density function for permeability is therefore vital for reliable simulations. A number of studies \cite{S,H,SS,SP,SPW,ESM,SMP} have used a normal probability density functions for experimentally determined  permeability at macroscale. But their measurements have been mostly for small sample sizes and hence may still be the subject to experimental and statistical inaccuracy. In other words, data obtained from the experiments may not be enough to choose between two (normal or lognormal distribution) or more competing distribution functions. Furthermore, these studies ignore the effect of microscale uncertainties on macroscale permeability uncertainties. \\
Different approximation methods have been used to determine the impact of microscale uncertainties on macroscale permeability uncertainties, e.g, finite element based Monte Carlo and Lattice-Boltzmann methods have been used to estimate permeability and superficial velocity of representative volume elements of porous media \cite{G,SSP,A,B,C}. The accuracy of the analytical methods has been debated because they have considered homogeneous periodic arrays of parallel fibres instead of random distributions. In addition, all of the modeling approaches require that either the distribution of at least one property of fibrous media or the distribution of macropors of fibrous media be known. Another criticism is that estimating permeability by curve fitting with empirical constants is known to generate significant systematic errors. \\
In view of the fact that finding an appropriate distribution function describing the spatial variation of permeability in fibrous media is a challenging problem. Therefore, a method that measures the microstructural variability as input for stochastic simulations is required. In the \cite{GBN}, we analyzed the effect of tortuosity on the variability of permeability at the average local fibre volume fraction(microscale level). We showed that the Gaussian distribution is not necessarily the most appropriate distribution for representing permeability \cite{GBN}. In this study, we capture the influence of a distribution of local fibre volume fraction ($V_f$) on the variability of permeability. The uncertainty in this variable (e.g,$V_f$) propagates to a larger level and is reflected in the variability of the geometry of flow affecting the final quality of composites.
In order to establish a probability density function for permeability, in this study we propose that (i) the best probability density function may be approximated for distribution of fibre volume fraction by a normal model, (ii) the values of fibre volume fraction were used to compute the distribution of permeability applying the Kozeny-Carmen equation, (iii) we used the Kozeny-Carmen equation together with change of variable technique to determine the probability density function of permeability and subsequently the analytical approach is compared with the distribution of permeability \cite{F} (Figure~\ref{fig:flowchart1}).

\section{The probability density function (PDF) of $K$}
\begin{figure}[h]
  \centering
  \includegraphics[scale=0.5]{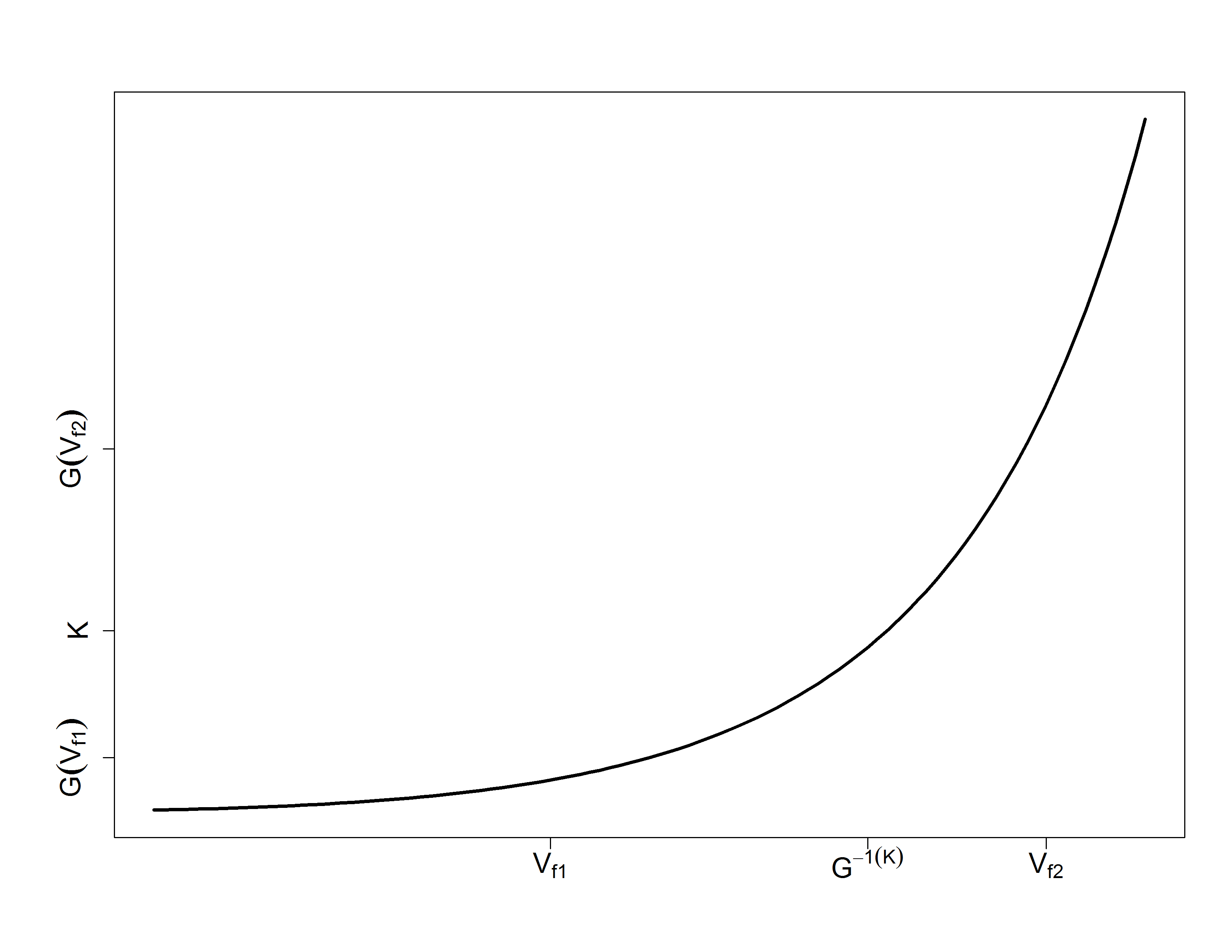}
  \caption{Continuous variation of K values as function of $V_f$}
  \label{change}
\end{figure}

 There is no  universal established relationship between fibre volume fraction and permeability. In this paper we recall the Carmen Kozeny equation in order to find the PDF of permeability (\ref{eq.Kozeny}).
Observe that in (\ref{eq.Kozeny}) the random variable (RV) $K$ is an increasing function of porosity $V_f$ (assuming $r_f^2$, $k_c$ are constant), see Figure~\ref{change}. Note also that $K$ is significantly affected by $r_f^2$. However this will not be addressed in this paper.  Here we use change-of-variables technique, in this study called "the change of $V_f$", to investigate PDFs of permeability. This technique is a common and well-known way of finding the PDF of $y=y(x)$ if $x$ be a continuous random variable with a probability density function $f(x)$.  Therefore to establish the PDF of  random variable $K$, it is required to have the PDF of $V_f$. Thus subsequently in the next section, we determine numerically the PDF of $V_f$.

\subsection{Fibre volume fraction ($V_f$)}
\def\bbE{\mathbb{E}}
 Variable fibre volume fraction for a fibrous media depends on areal weight density ($A_w$) and thickness(t) and can be expressed as equation (\ref{eq.poro}) :
\beq \label{eq.poro}
V_f=\frac {{A_w}\; n} {{\rho_f }\;t}
\eeq
where $\rho_f$ stands the fibre density, n is number of layers. To attain a closed form expression of the density of ${V_f}$, both ${A_w}$ and ${t}$ are considered normal random variables with correlation coefficient, -1<$\rho_{corr}$<1. When $\rho_{corr}$=0, the two variables ${A_w}$ and ${t}$ are independent, the distribution of ${V_f}$ would have a Cauchy distribution. Note that the the Cauchy distribution does not have finite moments of any order hence the mean and variance of $V_f$ are undefined. Therefore, assuming a Cauchy distribution would not be an appropriate model for fibre volume fraction.
Now set $\mu_{A_w}:=\bbE(A_w)$ and $\mu_t:=\bbE(t)$. In this stage of work we consider the first order Taylor expansion about $({\mu_{A_w}},{\mu_t})$ for $V_f({A_w},{t}$):
\beq\label{eq.Taylor}
V_f:=V_f(A_w,t)=V_f(\mu_{A_w},\mu_t)+V'_{f_{A_w}}(\mu_{A_w},\mu_t)({A_w}-\mu_{A_w})+V'_{f_t}(\mu_{A_w},\mu_t)(t-\mu_t)+O(n^{-1}),\eeq
where $V'_{f_{A_w}}$ and $V'_{f_t}$ are the derivatives of $V_f$ with respect to ${A_w}$ and $t$ respectively. In agreement with \cite{SO,EJJ}, the approximation for $\mu_{V_f}:=\bbE(V_f)$ is given by
\beq\label{exp:V} \begin{array}{ccl}
\mu_{V_f}&=&\bbE\Big(V_f(\mu_{A_w},\mu_t)+V'_{f_{A_w}}(\mu_\rho,\mu_t)({A_w}-\mu_{A_w})+V'_{f_t}(\mu_{A_w},\mu_t)(t-\mu_t)+O(n^{-1})\Big)\\
&\approx&\bbE\big(V_f(\mu_{A_w},\mu_t)\big)+V'_{f_{A_w}}(\mu_{A_w},\mu_t)\bbE({A_w}-\mu_{A_w})+V'_{f_t}(\mu_{A_w},\mu_t)\bbE(t-\mu_t)\\
&=&\bbE\big(V_f(\mu_{A_w},\mu_t)\big)={n\;{\mu_{A_w}}}\Big/ {{{A_w}_f }\;\mu_t}.\end{array}\eeq
By virtue of the definition of variance we can write
 \beq \label{eq.var} \sigma^2(V_f)=\sigma^2\Big(\frac{{{A_w}}}{{t}}\Big)\Big(\frac{n}{{A_w}_f }\Big)^2.\eeq
Next, use the first order Taylor expansion once again around $({\mu_{A_w}},{\mu_t})$. Then owing to (\ref{exp:V}) we approximate
\beq\label{eq:apx.V} \sigma^2(V_f)\approx\bbE\Big\{\big(V_f({A_w},t)-V_f(\mu_{A_w},\mu_t)\big)^2\Big\}.\eeq
Substitute (\ref{eq.Taylor}) in (\ref{eq:apx.V}), then (\ref{eq.var}) becomes the following:
\beq \label{eq.var.1}
\begin{split}
 \sigma^2(V_f)&\approx\Big(\frac{n}{{A_w}_f}\Big)^2 \sigma^2\Big(\frac{\mu_{A_w}}{\mu_h}+\frac{1}{\mu_h}(\rho-\mu_\rho)-\frac{\mu_\rho}{\mu^2_h}(h-\mu_h)\Big)\\
&=\Big(\frac{n}{\rho_f}\Big)^2\sigma^2\Big(\frac{1}{\mu_h}\rho-\frac{\mu_{A_w}}{\mu^2_t}t\Big)\\
&=\Big(\frac{n}{{A_w}_f}\Big)^2\frac{1}{\mu^2_t}\;\sigma^2({A_w})+\frac{\mu^2_{A_w}}{\mu^4_t}\sigma^2(t)-2\frac{\mu_{A_w}}{\mu^3_h}Cov({A_w},t)\\
&=\Big(\frac{n}{{A_w}_f}\Big)^2\frac{\mu^2_{A_w}}{\mu^2_t}\;\bigg(\frac{1}{\mu^2_{A_w}}\sigma^2({A_w})+\frac{1}{\mu^2_t}\sigma^2(t)-2\frac{1}
{\mu_{A_w}\mu_t}Cov({A_w},t)\bigg)
\end{split}
\eeq

In (\ref{eq.var.1}), $Cov({{A_w}},{t})$ expresses the covariance of ${{A_w}},\;{t}$ while as we said ${\mu_{A_w}}$ and ${\mu_t}$ are average of local areal density and  thickness, respectively.

\begin{figure}[h!]
\centering
\begin{minipage}{.50\linewidth}
\centering
\subfloat[]{\label{main:a}\includegraphics[width=0.95\linewidth]{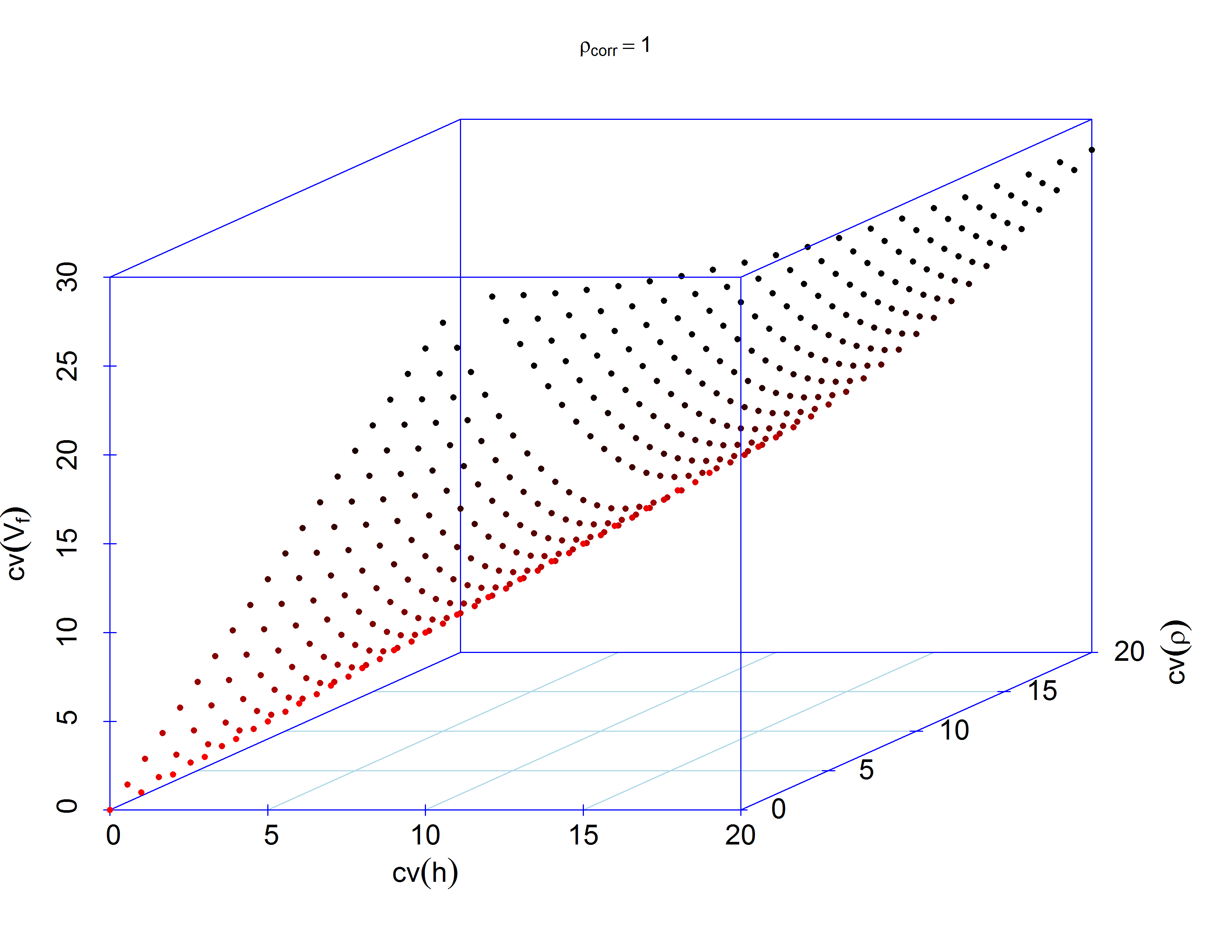}}
\end{minipage}%
\centering
\begin{minipage}{.50\linewidth}
\centering
\subfloat[]{\label{main:a}\includegraphics[width=0.95\linewidth]{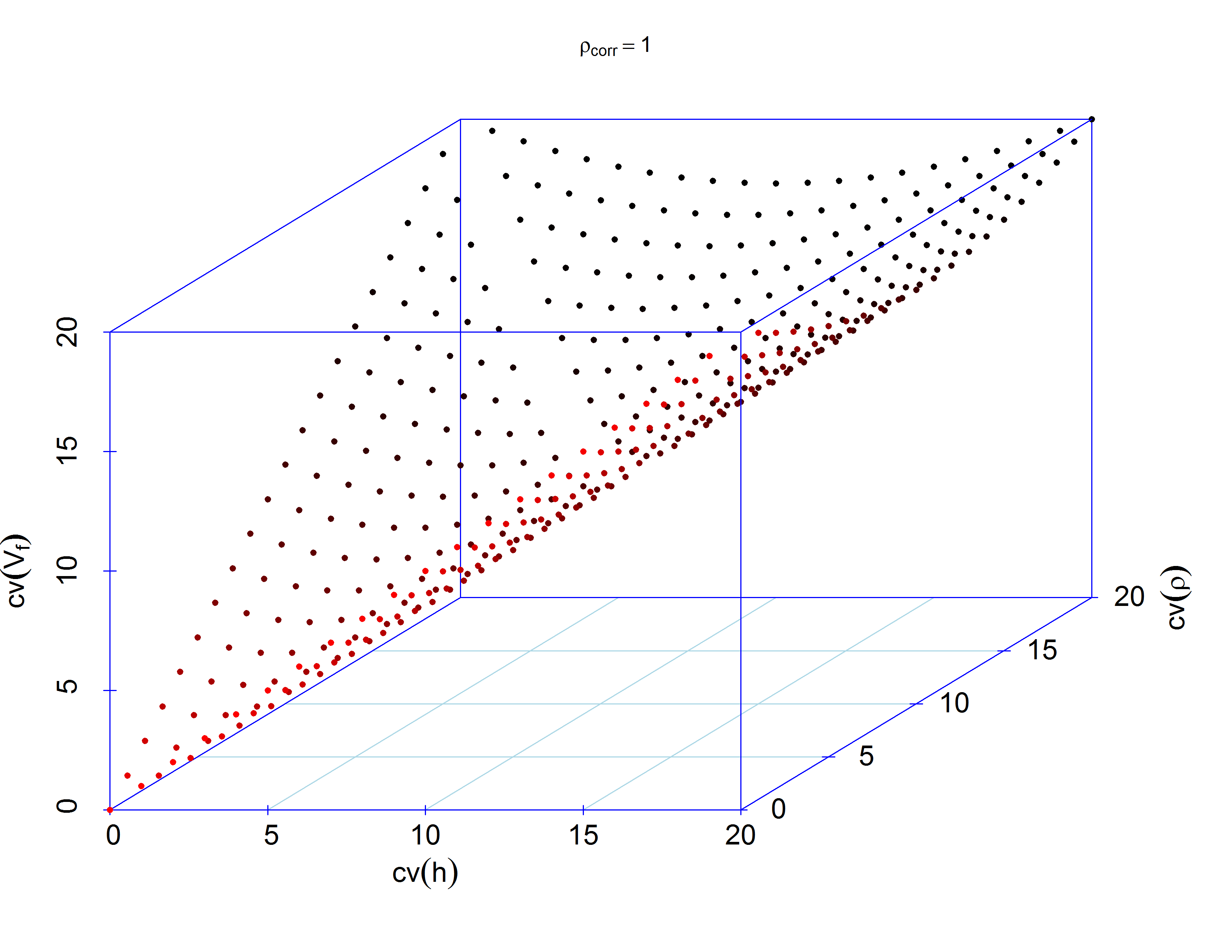}}
\end{minipage}%
\par\medskip
\begin{minipage}{.50\linewidth}
\centering
\subfloat[]{\label{main:a}\includegraphics[width=0.95\linewidth]{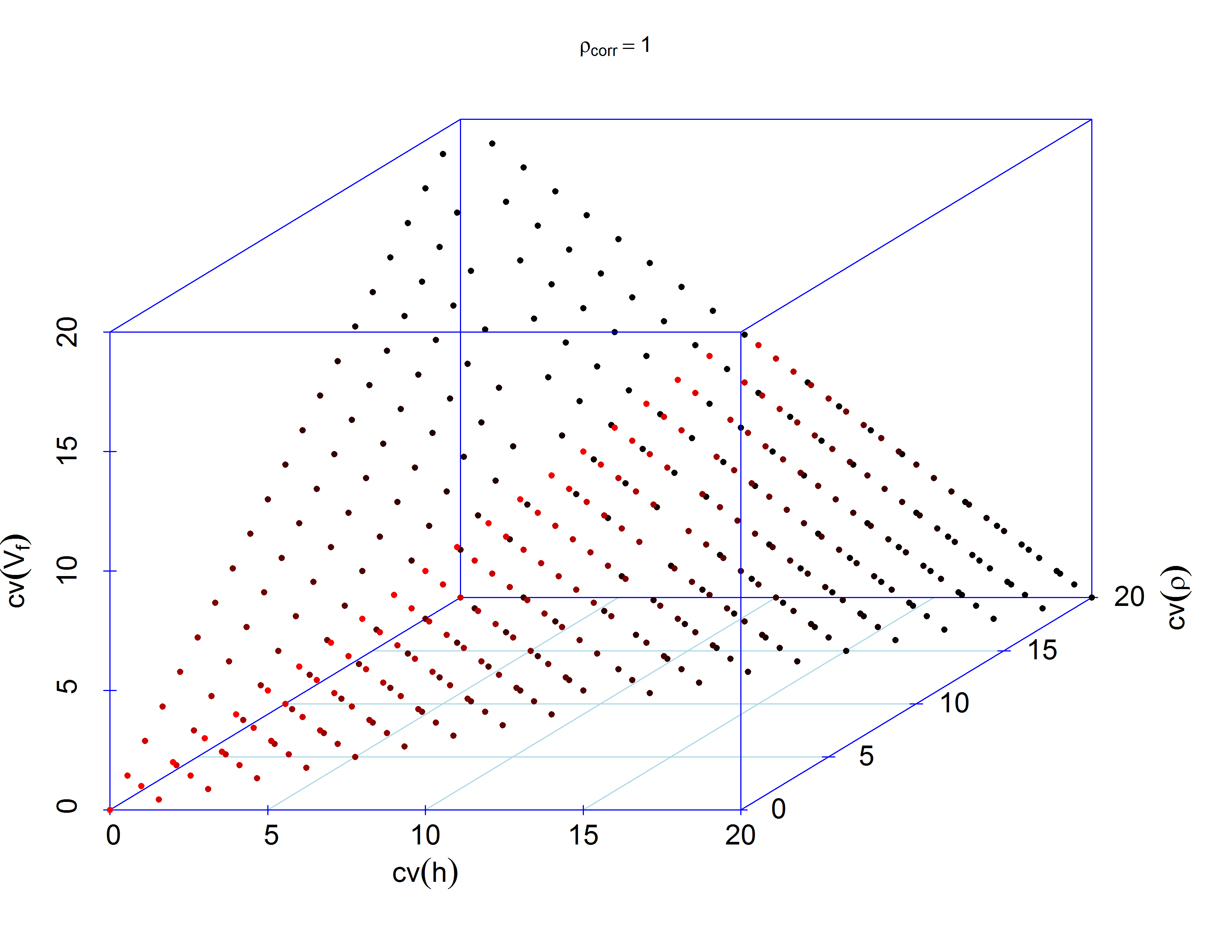}}
\end{minipage}%
\caption{Coefficient of variation of fibre volume fraction $cv({V_f})$ as a function of coeffiecient of variation of areal density,$cv({{A_w}})$, and thickness, $cv({t})$, at ${\mu_{V_f}}=0.5$.(a)$\rho_{corr}=0$, (b)$\rho_{corr}=0.5$, (c)$\rho_{corr}=1$}
\label{fig:hhh}
\end{figure}

Now denote $\sigma({{A_w}})$, $\sigma({t})$ as the standard deviation of RVs ${A_w}$, $t$ and moreover $\rho_{corr}$ as their correlation coefficient and their relationship can be expressed as:
\beq \label{eq.stan} Cov({{A_w}},{t})=\sigma({{A_w}})\sigma({t})\rho_{corr}.\eeq
In addition we define the coefficient of variation ($cv$) of a RV, such as  $cv({V_f})$:

\beq  cv({V_f})=\frac{\sigma({V_f})}{\mu_{V_f}}
\label{eq.cvvv}
\eeq
Consequently  Eqn. (\ref{eq.var.1}) can be recast as:
\beq  \sigma^2({V_f})\approx(\mu_{V_f})^2(cv^2({{A_w}})-2cv({{A_w}})cv({t}){A_w}_{corr}+cv^2({t}))\label{eq.var.2}\eeq
Consider (\ref{eq.var.2}) and replace it in (\ref{eq.cvvv}), then the $cv$ of fibre volume fraction yields:
\beq cv({V_f})\approx\sqrt{cv^2({{A_w}})-2cv({{A_w}})cv({t})\rho_{corr}+cv^2({t})}
\label{eq.var.3}\eeq
Equation (\ref{eq.var.3}) shows that the $cv$ of RV fibre volume fraction${\V_f}$ is approximately independent from $\mu_{V_f}$. Owing to (\ref{eq.var.3}) at point $\mu_{V_f}=0.5$, evidently 3D Scatter plots (\ref{fig:hhh}) exhibits  variation of  the $cv$ of the RV fibre volume fraction ${\V_f}$ as a function of the coefficient of variations thickness $cv({t})$ and local areal density $cv({{A_w}})$.\\
In Figure \ref{fig:hhh}, it is possible to observe that when $\rho_{corr}$ increases from zero to one, a saddel is formed with $\rho_{corr}=1$, $cv({{A_w}})=cv({t})$ and then $cv({V_f})=0$. Furthermore, Figure \ref{fig:hhh} shows that as $cv({{A_w}})$ and $cv({t})$ approach each other too closely, $cv({V_f})$ moves from the top to the lowest level.\\
As to the PDF of $V_f$, section 2.1.1 establises experimentally that the random variable $V_f$ has normal distribution. Although by calling the following cases we claim that this assertion in independent case is not theoretically artificial:
\begin{itemize}
\item[1.] ${A_w}$ and $t$ are {\bf independent}. It is straightforward that when a random variable $t$ follows Cauchy distribution with parameter $\gamma$ then $1/t$ has also Cauchy distribution with parameter $1/\gamma$. Further, according to our explanation above, we can easily check that the ratio of two independent normal random variables determines the Cauchy PDF. Hence as result if we consider random variables ${A_w}$ and $t$ takes normal and Cauchy distributions respectively then $V_f$ has normal PDF.
\item[2.] ${A_w}$ and $t$ are {\bf dependent}. This case is more complicated but practical. First note that if one is interested in bivariate normal distribution for pair $({A_w},t)$. We address to \cite{Hi} which study the ratio of two correlated normal random variables. The author indeed has established that if ${A_w}$ and $t$ be normally distributed random variables with means $\mu_i$, variances $\sigma^2_i$, $(i=1,2)$ and correlated coefficient $\rho_{corr}$, then the exact PDF of $V_f$ takes the form (1) in page 636 in \cite{Hi}. For simplicity we omit the form (1) here. However, in \cite{CKB} we could observe that in this special case, i.e. normally distributed $({A_w},t)$ the PDF of $V_f$ is not necessarily symmetric and normal. Therefore by virtue of author's investigations, it is not clear yet that what kind of distributions should be considered for $({A_w},t)$ to prove analytically a normal PDF for $V_f$. Consequently, the solution which may come cross the mind is experimental results represented in next subsection.
\end{itemize}

\subsubsection{Experimental}
Table~\ref{input} shows the specifications of a $2\times2$ twill carbon woven fabric used for production of composite parts in this study. The Fabric  was cut in the warp direction. Composite parts were produced with same ${\V_f}$ using High Injection Pressure Resin Transfer Moulding(HIPRTM). Full production details have been presented in \cite{CKB}.
\begin{table}[!htbp]
\caption{Material properties of fabrics used to prepare composite samples}
\begin{center}
    \scalebox{0.80}{\begin{tabular}{| l |  l | l | l | l |l | }
    \hline
    Style & Weave pattern &Areal weight$(gm^{-2})$ & Maximum width$(cm)$&Fiber  diameter$(\mu m)$& Number of filaments  \\ [0.3ex]
\hline
   280T & $2\times2$ twill& 280&100&7&3000\\ [0.3ex]
    \hline
   \end{tabular}}
\label{input}
\end{center}
\end{table}

\begin{figure}[htbp]
 \par\medskip
\centering
\begin{minipage}{.92\linewidth}
\begin{center}
\subfloat[]{\label{main:a}\includegraphics[scale=.18]{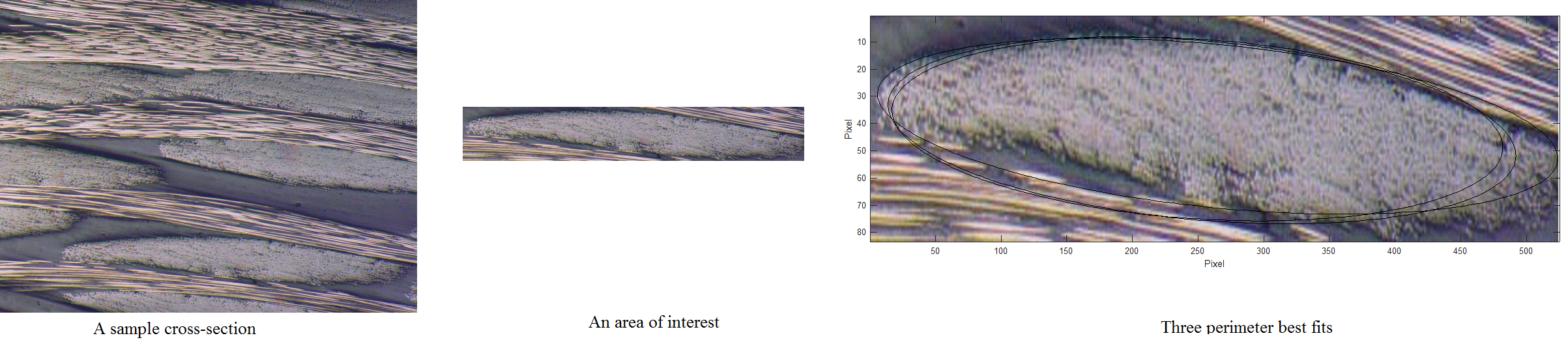}}
\end{center}
\end{minipage}%
\par\medskip
\begin{minipage}{.90\linewidth}
\begin{center}
\subfloat[]{\label{main:aa}\includegraphics[scale=.5]{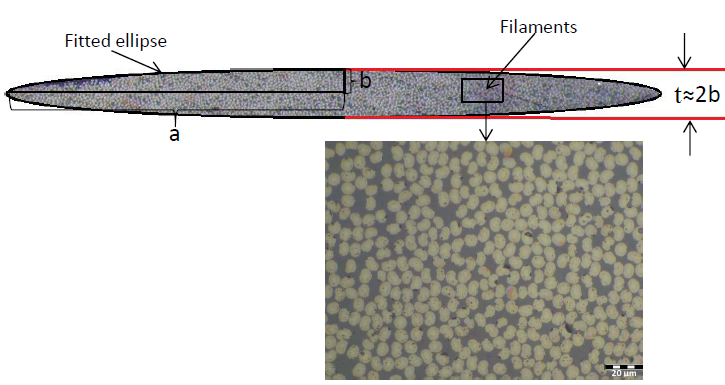}}
\end{center}
\end{minipage}%
\caption{(a) Ellipse detection, (b)Tow geometry parameters}
\label{ellipse}
\end{figure}

 In order to measure the h of each tow, series of samples were cut perpendicular to the fibre direction, the samples were  polished manually in four steps (sandpaper grits 320, 400, 800, and 1000), and subsequently photographed using an Olimpus PNG3 optical microscope equipped with a CCD camera. The analysis of each tow is carried out  by image processing Matlab \texttrademark 2015. Before importing the images into the  Matlab \texttrademark workspace, arears such as edges or borders which are not in the  interest of tow geometry characterizations were cropped. In projective geometry every tow section is equivallent to an ellipse. Figure \ref{ellipse}a  illustrates how an ellipse is fitted to a tow to locate the center point based on \cite{ellipse} and \cite{ellipse1}. Then the tow thickness ($t$) is equivalent to twice the length of the minor radius ($b$) (Figure \ref{ellipse}b). A total of 200 optical images were collected and an ellipse is fitted to each tow.


To determine the ${{A_w}}$  of each tow, the area of each tow (${A_t}$) in warp direction was approximated by the equivalent ellipse. Then, the number of filaments per tow (${n_t}$) were counted. Knowing the radius of the fibre cross section(r), length of warp tow (l), and ${\rho_f}$, the ${{A_w}}$  of each tow$({w_t})$ was computed from the following equation (\ref{eq.ad}):

\beq {{A_w}}=\frac{{n_t}\pi r^2 \rho_f}{{2a}}
\label{eq.ad}
\eeq


To determine distribution of ${V_f}$, the "R" statistical software \cite{R} was employed. For each pair (${t}$,${A_w}$), ${V_f}$ was computed, afterward a histogram was generated .
\subsubsection{Results}
\begin{figure}[h!]
\centering
\includegraphics[width=0.70\textwidth]{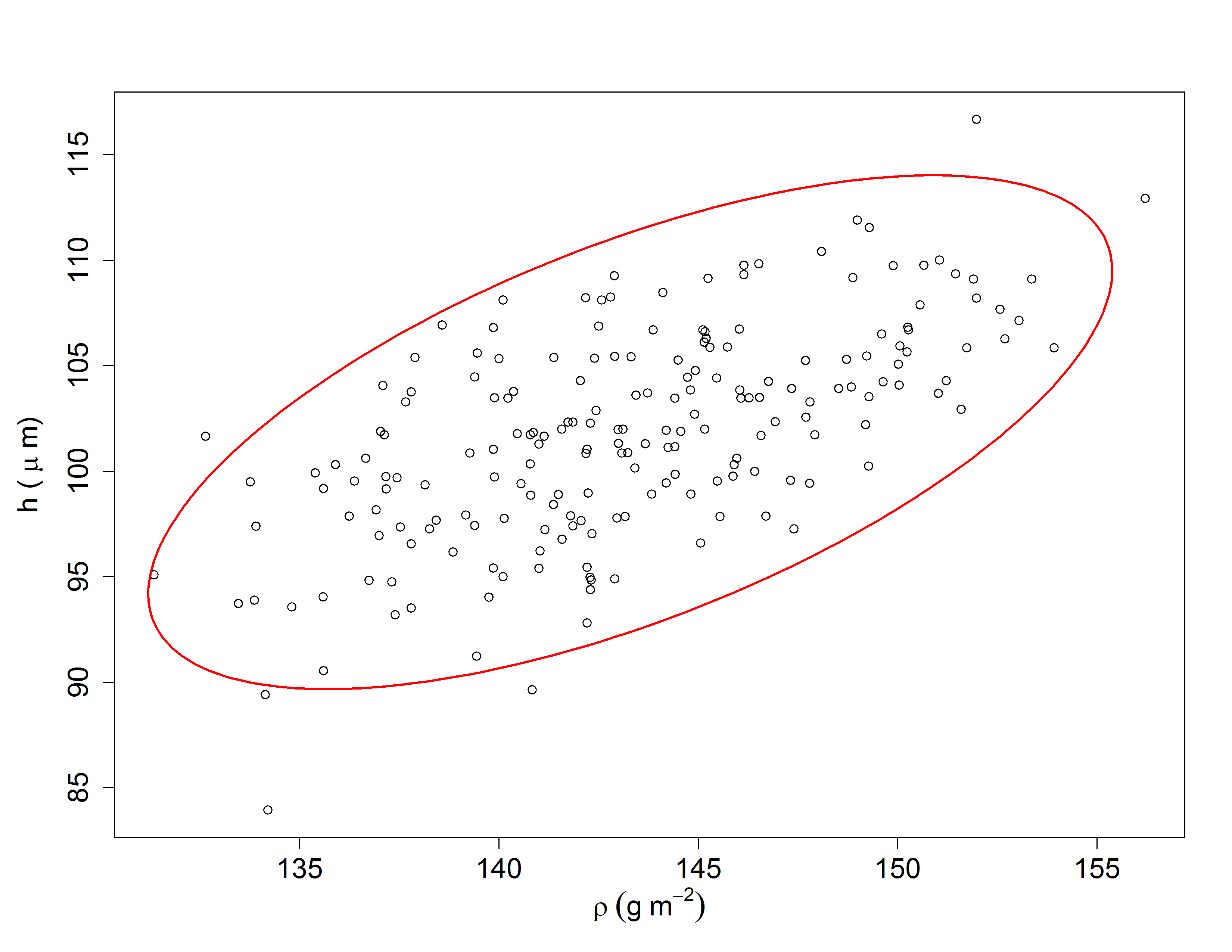}
		\caption{The relation of local average areal density with local average thickness. The bivariate normal ellipse (P=0.95) shows the data fit.}
	\label{fig:cv}
\end{figure}

A scatter plot of ${{A_w}}$ and ${t}$ for each tow is shown in  Figure \ref{fig:cv}. The coefficient of variation for the data was 0.72. An ellipse is fitted to the data. We can see that the ellipse extends between 100 and 200 $({g/m^2})$ on the 45 degree line. Hence ${{A_w}}$ and ${t}$ follow a bivariate normal distribution with mean components 168.5 ${g/m^2}$ and 102.3 ${\mu m}$, respectively. Therefore, our results suggest that  ${{A_w}}$ and ${t}$ are dependent and can be well approximated by the bivariate normal distribution.

\begin{figure}[h!]
\centering
\includegraphics[width=0.60\textwidth]{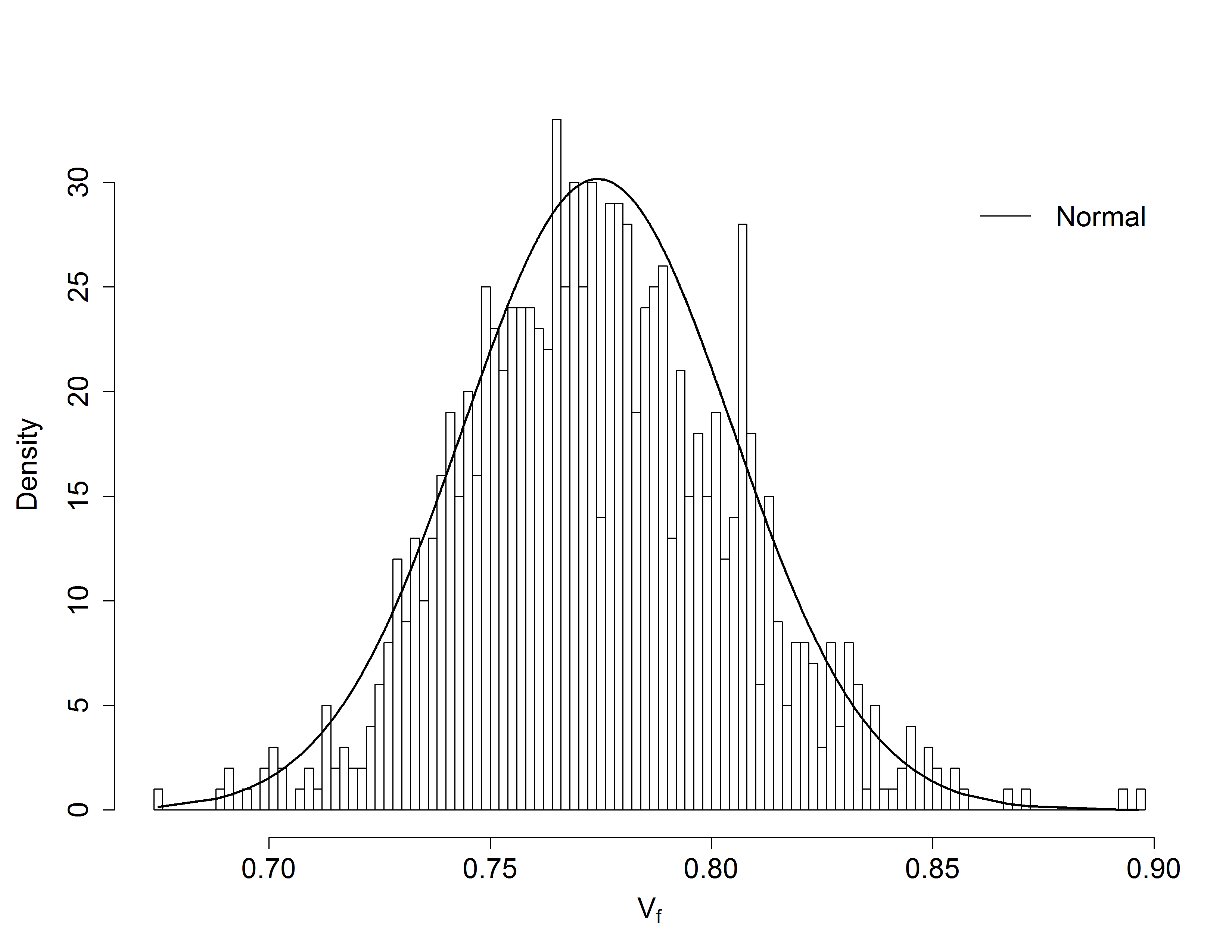}
		\caption{ Distribution of the local average fibre volume fraction.}
	\label{fig:hist1}
\end{figure}

\begin{figure}[h!]
\centering
\includegraphics[width=0.60\textwidth]{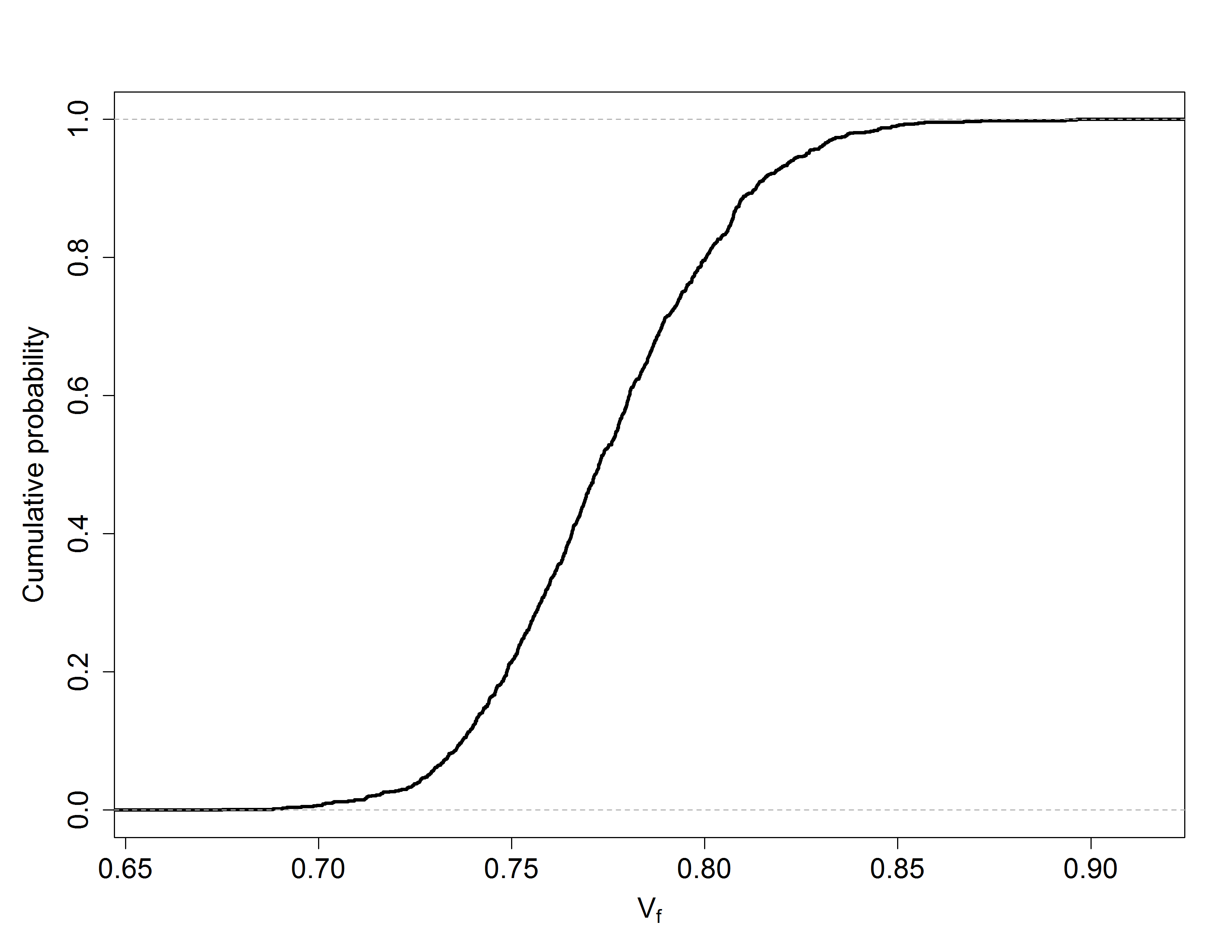}
		\caption{ Plot of CDF for local average fibre volume fraction}
	\label{fig:cdf}
\end{figure}

For each pair of ${{A_w}}$ and ${t}$,  ${V_f}$ was computed. Figure~\ref{fig:hist1} shows that the distribution of the local average fibre volume fraction follows a bell-curve distributions: the distribution of fibre volume fraction values are well approximated with a normal distribution model. A large distribution of ${V_f}$ was observed, ranging from $60 \%$ to $95\%$.
The graphical analysis indicates  the closeness of  the local average fibre volume fraction data  to  the  normal distribution  with cumulative distribution function(CDF)(Figure~\ref{fig:cdf}). The CDF plot is following a typical S curve indicative of normal distribution.
\subsection{ The PDF of Permeability ($K$)}
It follows from the section of $2.1$ that the fibre volume fraction of a preform, $V_f$, has a truncated normal PDF, $f^{{\rm TN}}$, with mean $\mu$ and variance $\sigma^2$ over $[V_{f1},V_{f2}]$ such that $0<V_{f1}<V_{f2}<1$:
\beq \label{PDF.TN} f^{{\rm TN}}(V_f)=\diy \frac{\diy\frac{1}{\sqrt{2\pi}}\exp\big(-\diy\frac{(V_f-\mu)^2}{2\sigma^2}\big)}
{\Phi\left(\diy\frac{V_{f2}-\mu}{\sigma}\right)-\Phi\left(\diy\frac{V_{f1}-\mu}{\sigma}\right)},\;\;\;\; V_{f1}<V_f<V_{f2}.\eeq
In Eqn. (\ref{PDF.TN}) and (\ref{PDF.TN1}) below, $\Phi(z)$ introduces the distribution function of a standard normal RV $Z$, that is
\beq \label{PDF.TN1}\Phi(z)=\mathcal{P}(Z\leq z)=\diy\int_{-\infty}^z \frac{1}{\sqrt{2\pi}}\exp\big(\frac{t^2}{2}\big)\rd t. \eeq
Passing to the limits $\V_{f1}\rightarrow 0$ and $\V_{f2}\rightarrow 1$ Eqn. (\ref{PDF.TN}) yields  the PDF of normal distribution with mean $\mu$ and variance $\sigma^2$.

Now assume the ratio $\diy\frac{r_f^2}{k_c}$ is constant, the RV $K$ is given by an increasing function of $V_f$, $K=G(V_f)$. Set $G^{-1}$ as inverse function of $G^{-1}$, i.e. $V_f=G^{-1}(k)$. \\

The Figure \ref{change} presents a graphical representation of function $G$ where $K$ is a monotonically increasing function of $V_f$.\\
Next, let $F_K$ be the distribution function of RV $K$, more precisely $F_K(k)=\mathcal{P}(K\leq k)$. Therefore, recalling $K=G(V_f)$ one can do:
\beq F_K(k)=\mathcal{P}(G(V_f)\leq k)=\mathcal{P}(V_f\leq G^{-1}(k))=F_{V_f}(G^{-1}(k)).\eeq
Where $F_{V_f}$ is the distribution function of $V_f$. Note that the second probability derives from the  of that $G$ is an increasing function of $V_f$.\\
Consequently by taking the derivative of $F_K$ with respect to $k$, the PDF of RV $K$ is obtained by
\beq \begin{array}{l} f_{K}(k)=\diy \frac{\rd}{\rd k}F_k(k)=\frac{\rd}{\rd k} F_{V_f}(G^{-1}(k))\\
\qquad =\diy \frac{\rd}{\rd k}G^{-1}(k). f_{V_f}(G^{-1}(k)). \end{array} \eeq

Introduce $k_1=G(V_{f1})$ and $k_2=G(V_{f2})$, since $G$ is an increasing function then $0<k_1<k_2$. Coming back to the assumption says that the RV $V_f$ has PDF $f^{{\rm TN}}(V_f)$ in $[V_{f1},V_{f2}]$, we obtain
\beq\label{eq.K1} \begin{array}{l} f_{K}(k)=\diy \frac{\rd}{\rd k}G^{-1}(k). f^{TN}_{V_f}(G^{-1}(k))\\
\quad =\diy \frac{\rd}{\rd k}G^{-1}(k).\frac{\diy\frac{1}{\sqrt{2\pi}}\exp\big(-\diy\frac{(G^{-1}(k)-\mu)^2}{2\sigma^2}\big)}
{\Phi\left(\diy\frac{G^{-1}(V_{f2})-\mu}{\sigma}\right)-\Phi\left(\diy\frac{G^{-1}(V_{f1})-\mu}{\sigma}\right)},\;\;\; k_1<k<k_2.\end{array}\eeq
As Figure \ref{change} shows this function is a one to one function so it is invertible. However, while it may be possible to find a closed form solution to this inverse, we will begin by presenting a numerical solution to this problem.

\subsection{ Numerical results}
\subsubsection{ Statistics of permeability}
\begin{figure}[h!]
\centering
\includegraphics[width=0.80\textwidth]{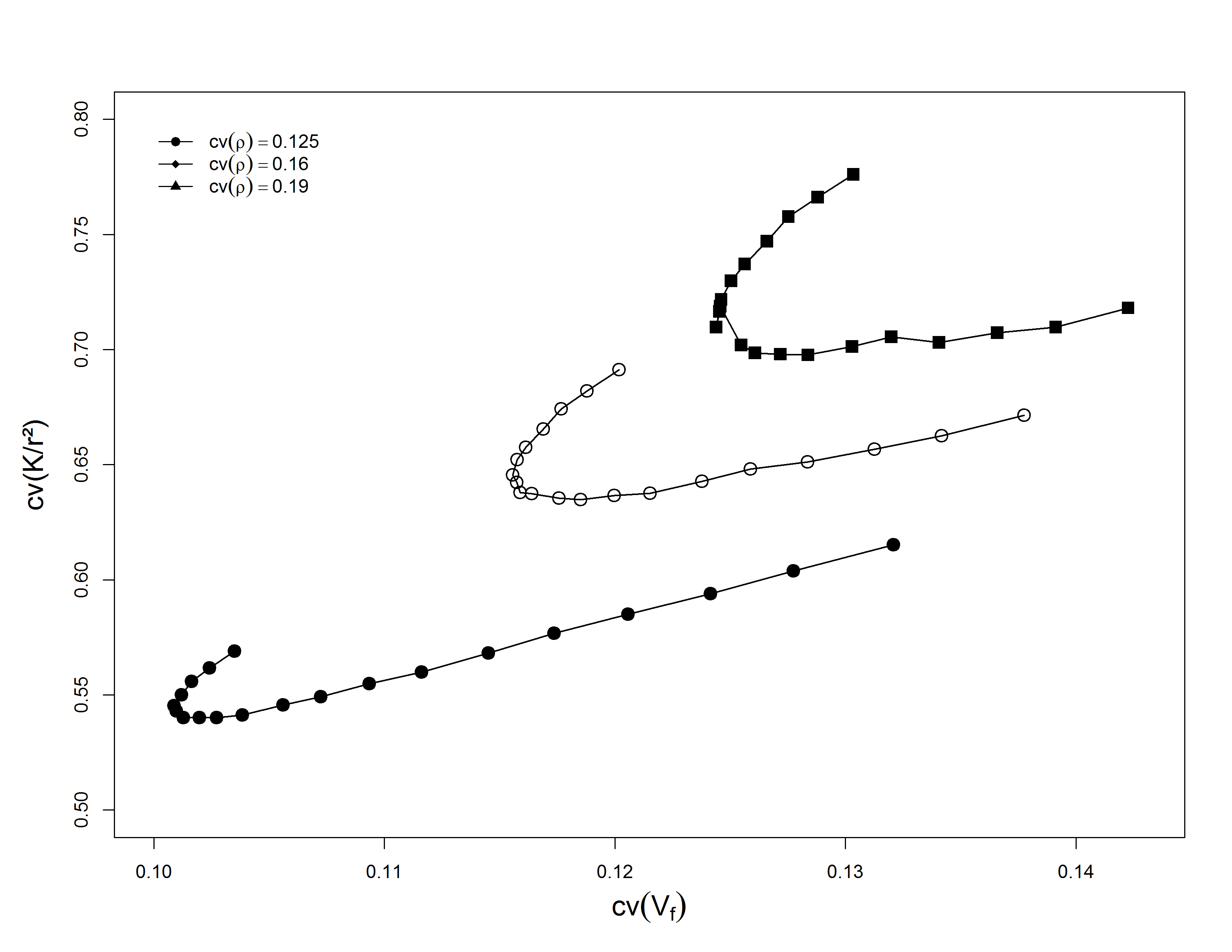}
		\caption{  the normalized local average permeability against coefficient of variation of local fibre volume fraction.}
	\label{fig:cvflow}
\end{figure}

 A summary of statistical properties of permeability for the different COV of $V_f $ is represented in Figure~\ref{fig:cvflow}. It is observed that an in-plane distribution of local fibre volume fraction results in a distribution of local permeabilities for flow perpendicular to the plane of the fibrous medium. It is established in Figure~\ref{fig:cvflow} that domain with larger local average areal densities possess larger local average permeability values. This phenomenon can be explained in terms of probability of number of contact points between fibres. The probability of number of contact points of fibres is larger for the domains with higher local average areal density. Higher contact points means stiffer arrangement of adjacent fibres, causing increased frictional resistance to fluid flow, and hence leading to less permeable area. Furthermore the normalized local average permeability (i) initially decreases when $cv({h})$ increases due to its influence on $cv({V_f})$, (ii) subsequently increases due to the increasing influence of $cv({h})$ on the normalized local average permeability. Therefore, as expected, this implies that the heterogeneity of fibrous media has a significant impact on the magnitude of variation in permeability.

\subsubsection{ Monte-Carlo simulation}
\begin{figure}[h!]
\centerline{\includegraphics[scale=0.5]{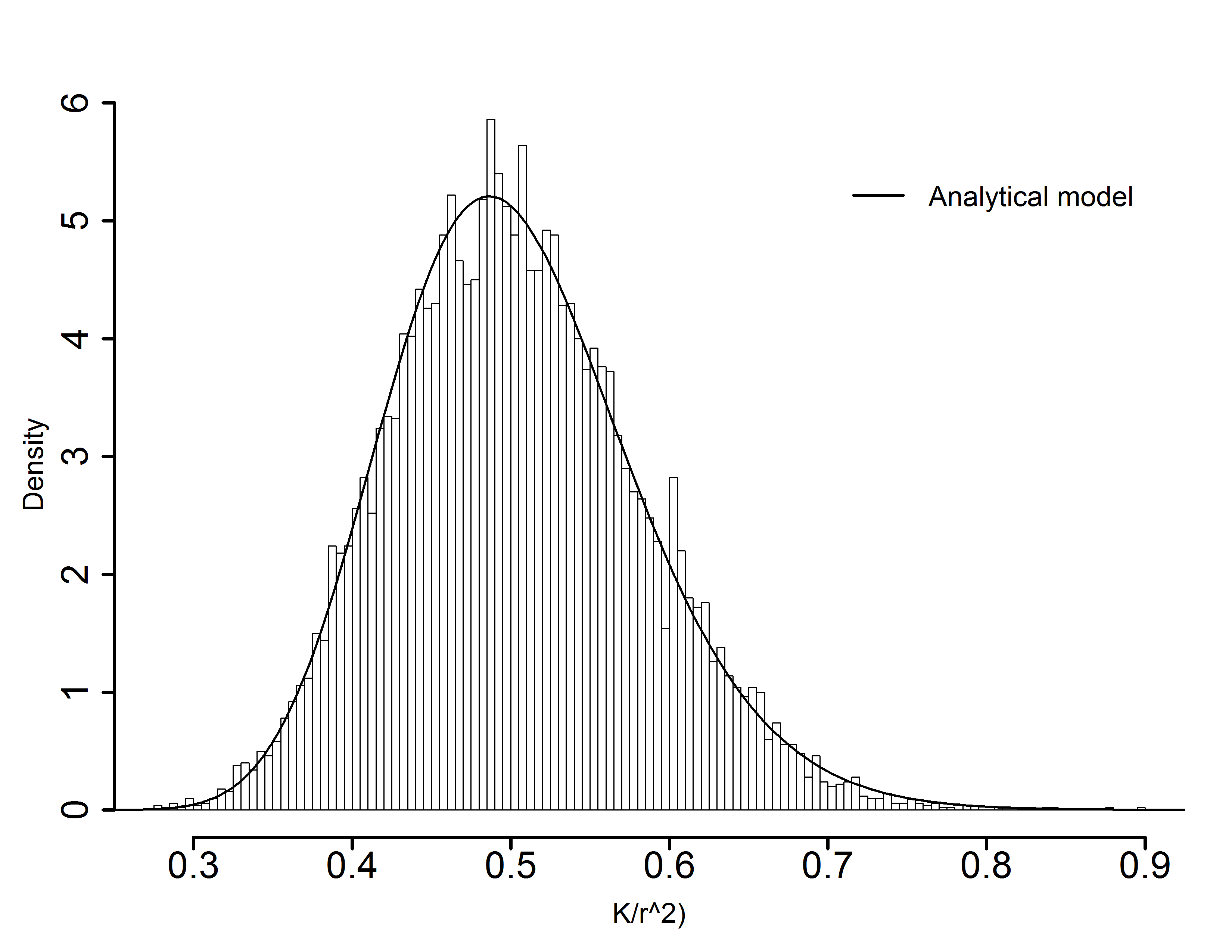}}
\caption{Fitting the computed PDF to histogram of the predicted permeability from Kozeny-
Carmen}
\label{MAP}
\end{figure}

The PDF (\ref{eq.K1}) is computed applying Monte-Carlos simulations (using the "R" statistical software \cite{R}). Monte-Carlo simulation was carried out on the domain with a average fibre volume fraction of 0.78. Subsequently this analytical model fit (\ref{eq.K1}) was compared with what would be predicted by physically based permeability equations (e.g. Kozeny-Carman(\ref{eq.Kozeny})). To do so, using the values of $V_f$ computed earlier, the distribution of permeability was obtained in (\ref{eq.Kozeny}). Then, probability values calculated through (\ref{eq.K1}) were fit to permeability values. Finally, in order to find the best distribution model to the analytical PDF, different statistical distributions are examined. \\
Figure~\ref{MAP} shows, obtained through the method described above, (i) the data are not symmetric and skewed to the right, (ii) there is significant agreement between the analytical approach and the simulation results of permeability (Kozeny-Carmen). The observed skewness in permeability distribution is close to what one was observed in \cite{par} at macrolevel.
The same principle was applied to various kinds of corrolations between $K$ and $V_f$. Table~\ref{com} lists the distribution behaviour of $K$ derived from models found in the literature.  The appearance of the distributions presented in table~\ref{com}  is the same as one shown in Figure~\ref{MAP}. Table~\ref{com} shows that the skewness of $K$ lies between 1.78-3.45, implying non-normal distribution of the permeability. In addition, the kurtosis values range between 9 and 24, deviating extremely from normality. As expected, there is a significant difference between the calculated skewness and kurtosis and that of the normal distribution.

\begin{table}[!htbp]
\caption{Comparison of distribution of normalized permeability  for different empirical equations}
\centering
\resizebox{\textwidth}{!}{\begin{tabular}{ |c|c|c|c|c|c|}
    \hline
   \multicolumn{1}{|c}{References} &\multicolumn{1}{|c|} {\(\displaystyle K/r^2\)} & \multicolumn{3}{|c|}{Goodness of fit}  \\ \cline{3-5}
     &  & Histogram & Skewness& Kurtosis \\

    \hline
 \Vcentre  {Gebart(Square) (1992)\cite{geba}} & \vbox{\begin{equation*} \frac{16}{9\pi \sqrt2}(\sqrt\frac{\pi}{4V_f}-1)^{2.5}\end{equation*}}\ & \raisebox{-\totalheight} {\includegraphics[scale=0.2]{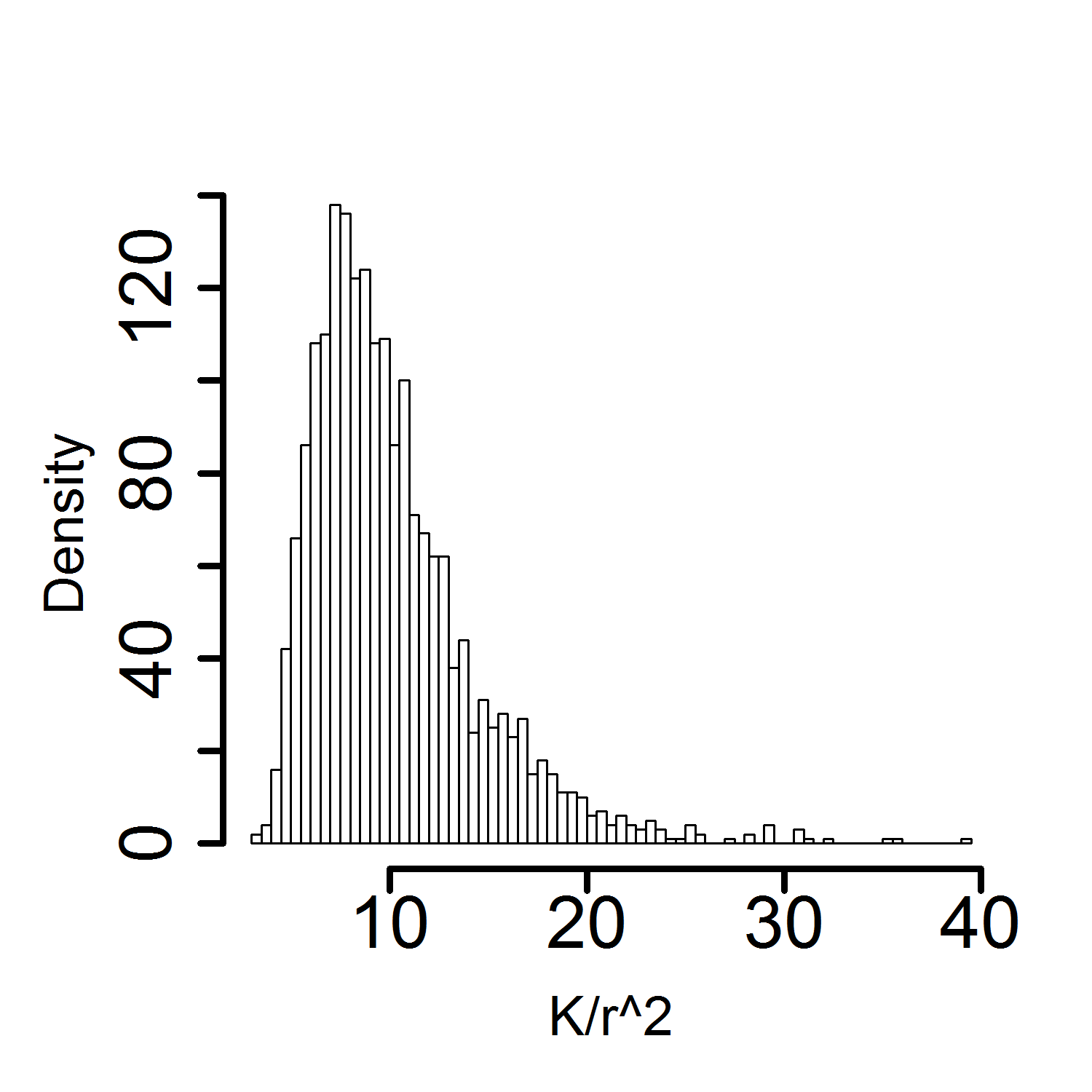}}\ & 2.22\ &12.22\ \\
    \hline
   Gebart(Hexagonal) (1992)\cite{geba} &\vbox{\begin{equation*} \frac{16}{9\pi \sqrt6}(\sqrt\frac{\pi}{2\sqrt3V_f}-1)^{2.5}\end{equation*}}\ &\raisebox{-\totalheight} {\includegraphics[scale=0.2]{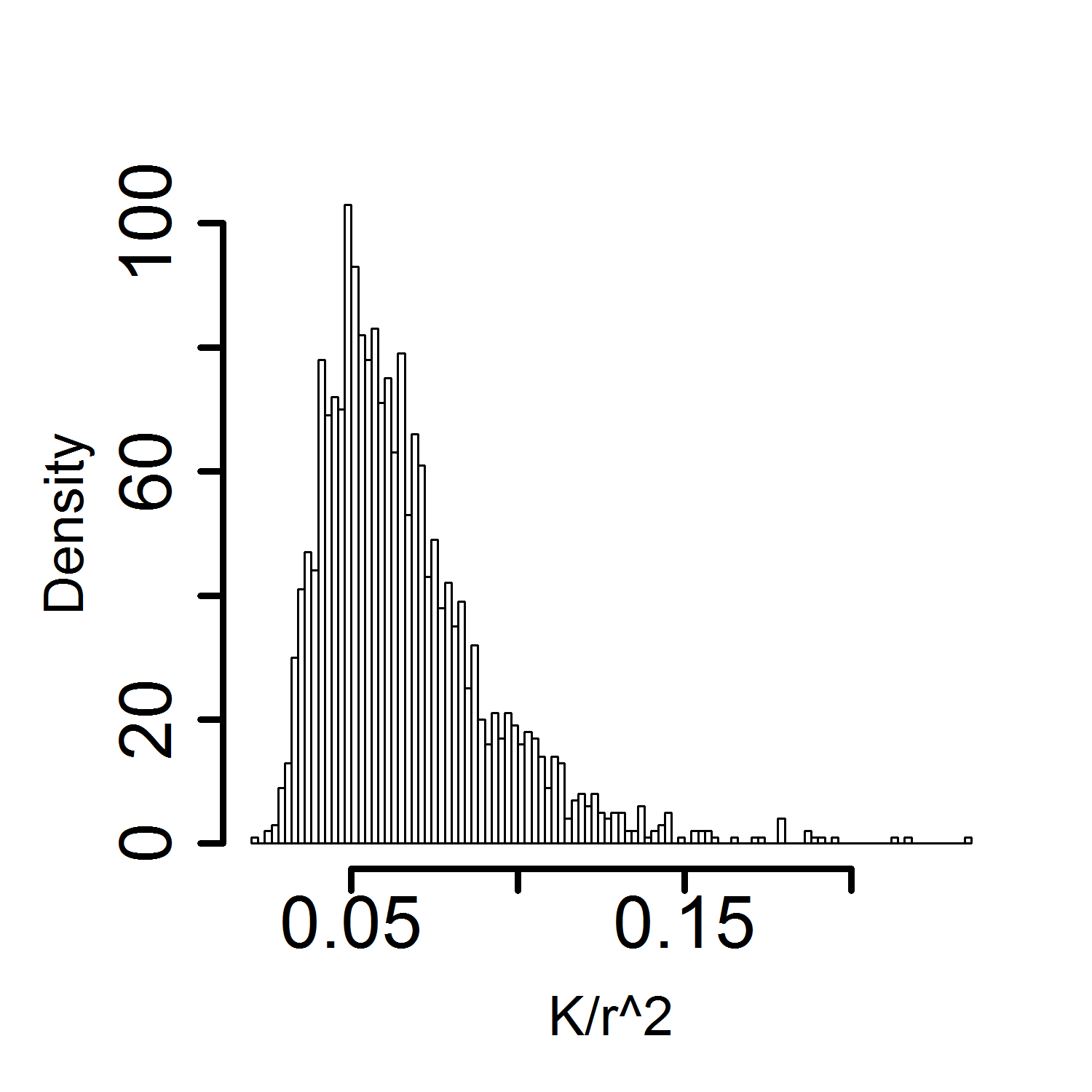}}\ & 2.12\ & 11.53\ \\
    \hline
   Bruschke and Advani (1993)\cite{BRU}& \vbox{\begin{equation*} \frac{1}{3}\frac{(1-L)^2}{L^3} (\frac{3L\arctan\sqrt\frac{1+L}{1-L}}{\sqrt{1-L^2}}+\frac{1}{L^2}+1)^{-1},\;\;\; \hbox{ (where {\(\displaystyle L^2=4V_f/ \pi\)} )}\end{equation*}}\ &\raisebox{-\totalheight} {\includegraphics[scale=0.2]{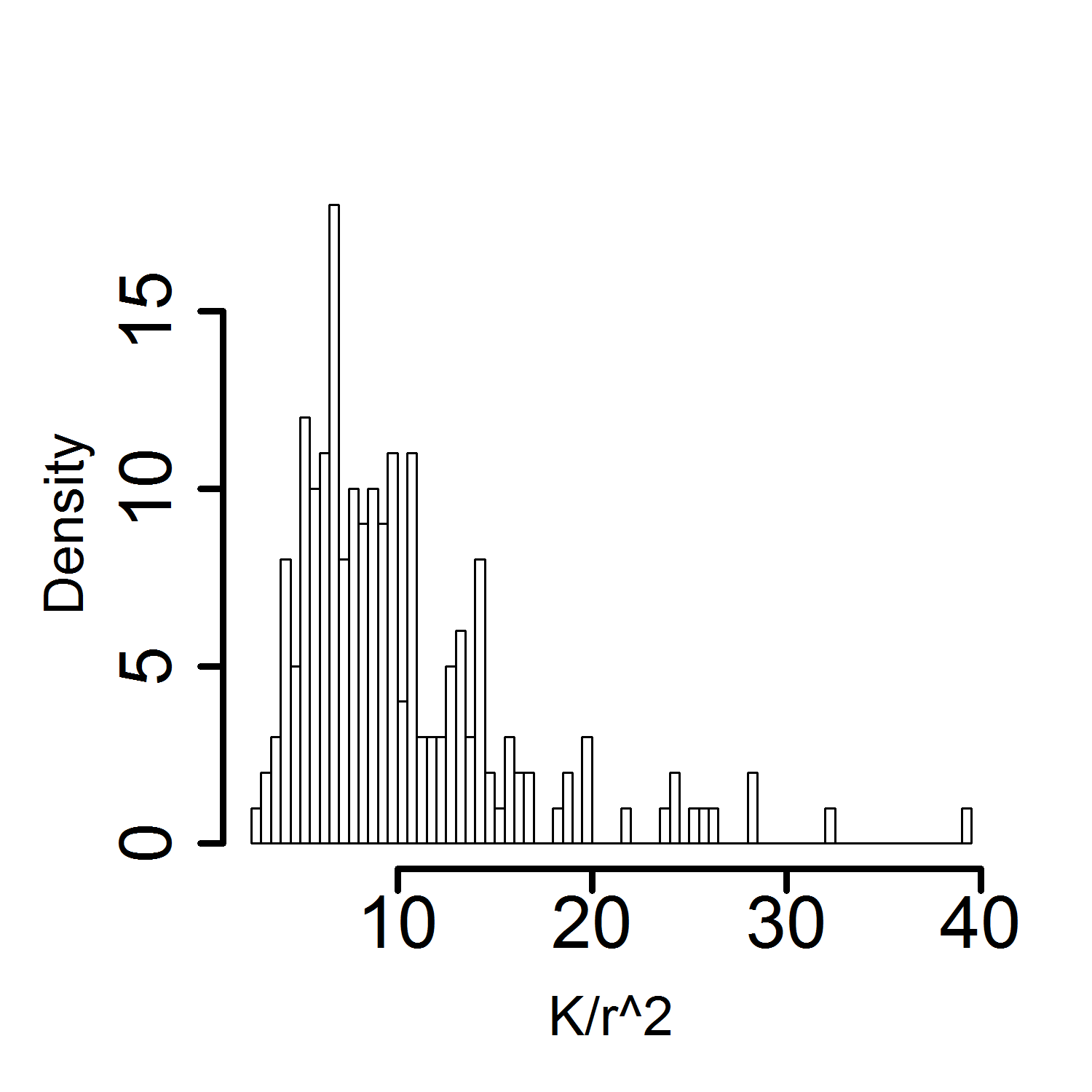}}\  &3.45\ & 24.15\  \\
 \hline
   Gutowski et al. (1987)\cite{GUT} & \vbox{\begin{equation*} \frac{1}{{V_f}^2} \frac{({{0.76}-{V_f}})^3}{{0.76}+{V_f}}\end{equation*}}\ & \raisebox{-\totalheight} {\includegraphics[scale=0.2]{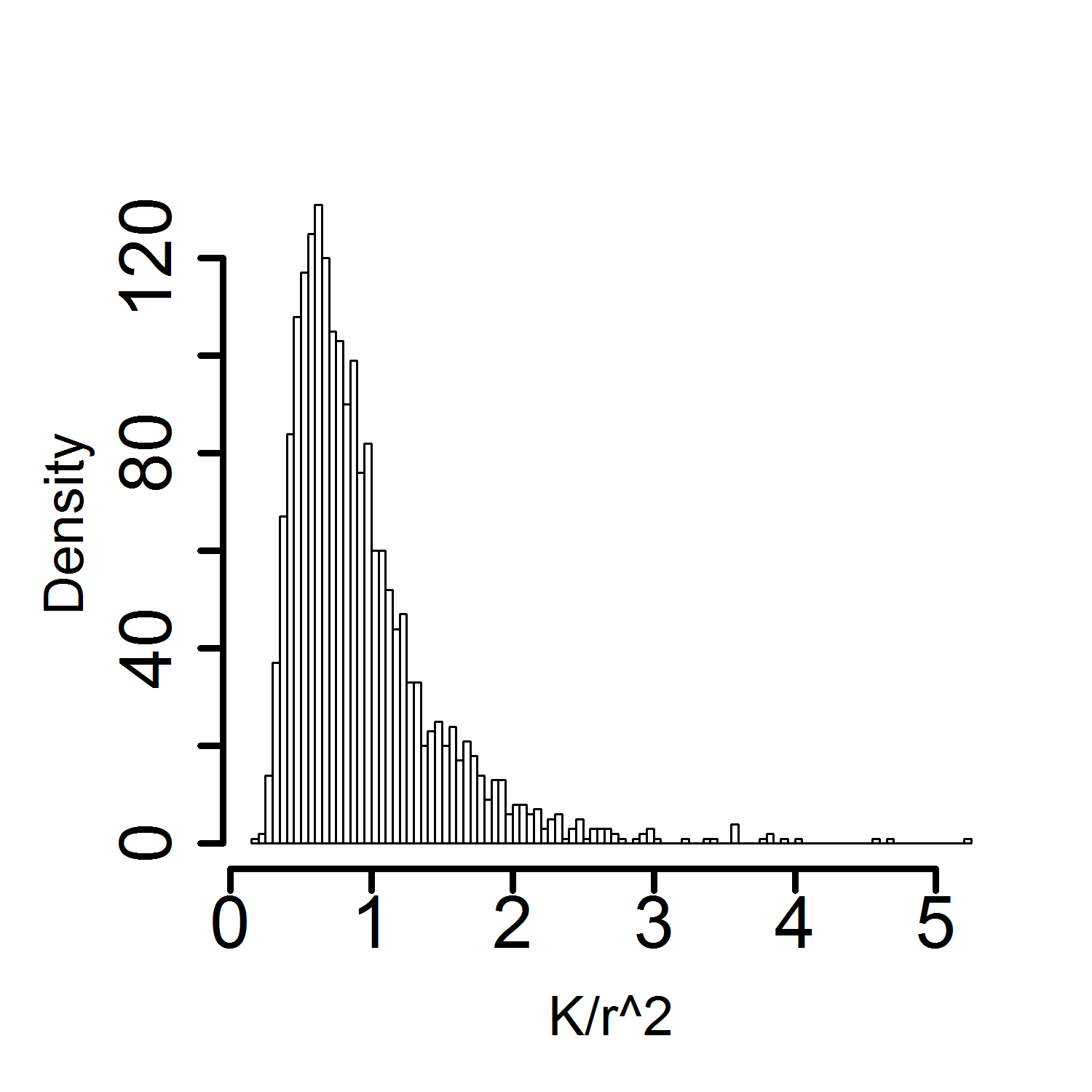}}\ &3.11 \ &20.38\ \\
 \hline
   Happel (1959)\cite{HAP}&\vbox{\begin{equation*} \frac{1}{9V_f} \frac{6-9{V_f}^{0.33}+9{V_f}^{1.67}-6{V_f}^2}{3+2{V_f}^{1.67}}\end{equation*}}\  &\raisebox{-\totalheight} {\includegraphics[scale=0.2]{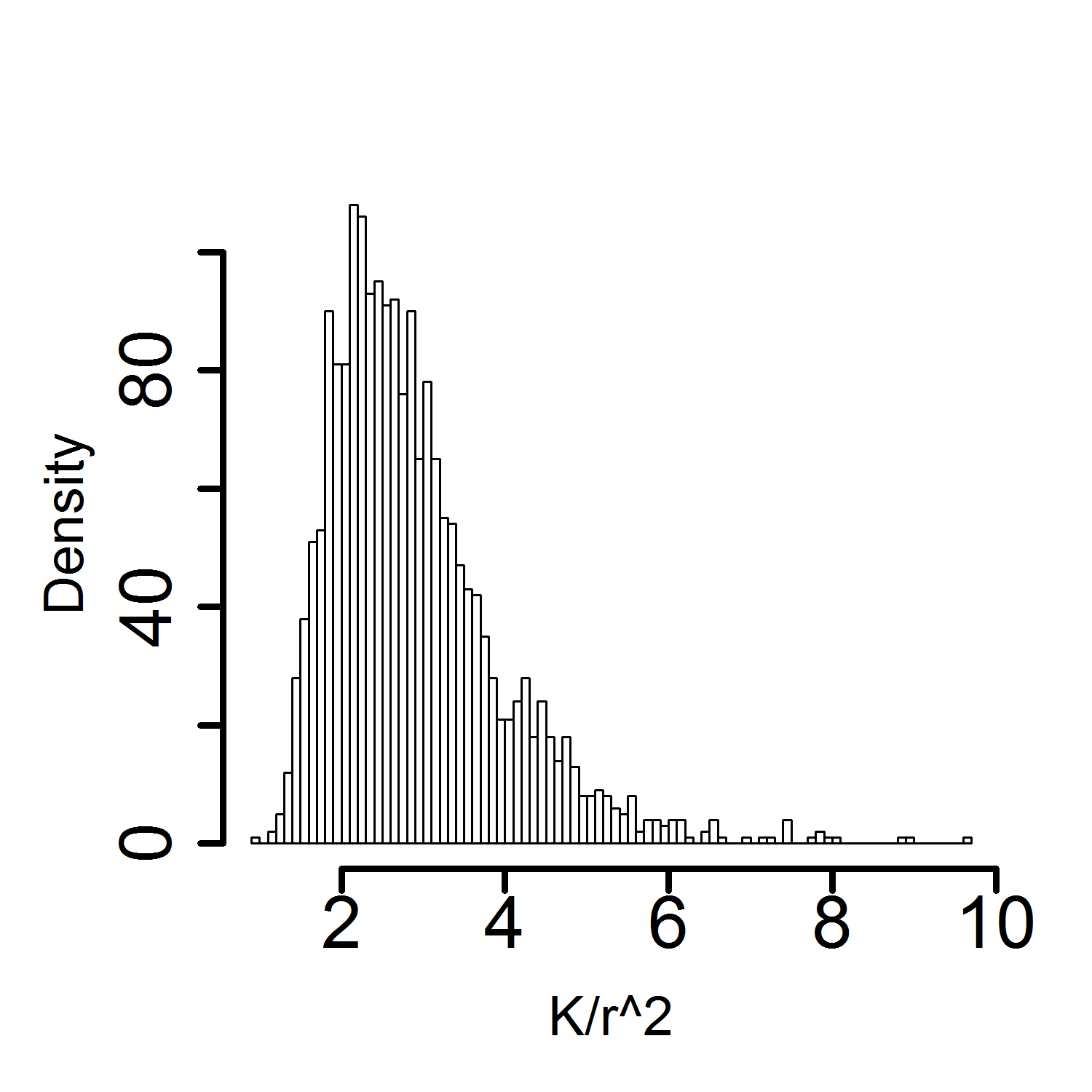}}\  &1.94\ &10.208\ \\
 \hline
  Lee and Yang (1997)\cite{lee}& \vbox{\begin{equation*}\frac{1}{8}\frac{(0.7854-V_f){(1-V_f)}}{{V_f}^{1.3}}\end{equation*}}\  &\raisebox{-\totalheight} {\includegraphics[scale=0.2]{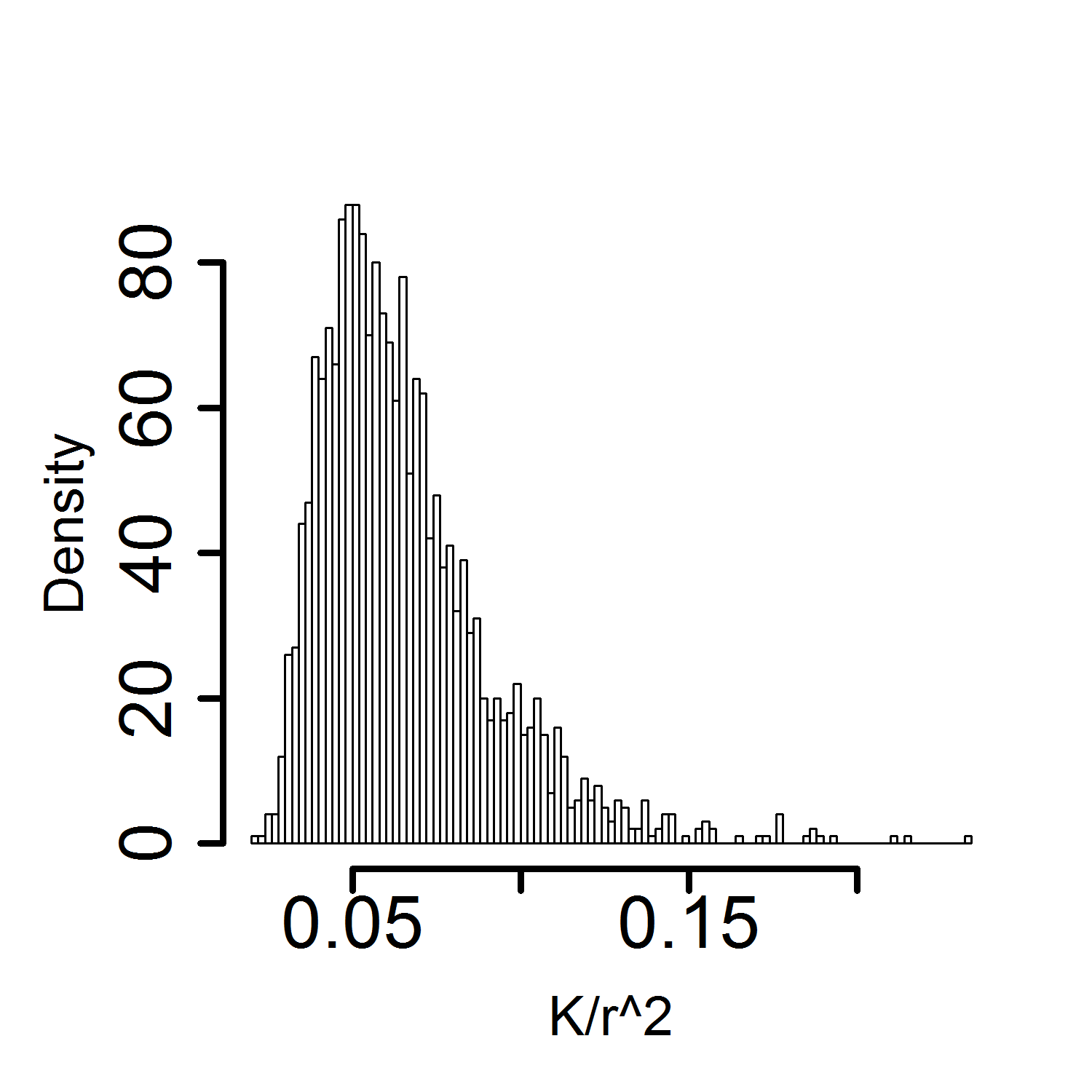}}\  & 2.01\ & 10.63\ \\
 \hline
 Sharaoui and Kaviany (1992)\cite{sha}&  \vbox{\begin{equation*}\frac{0.0602\pi (1-V_f)^{5.1}}{V_f}\end{equation*}}\   &   \raisebox{-\totalheight} {\includegraphics[scale=0.2]{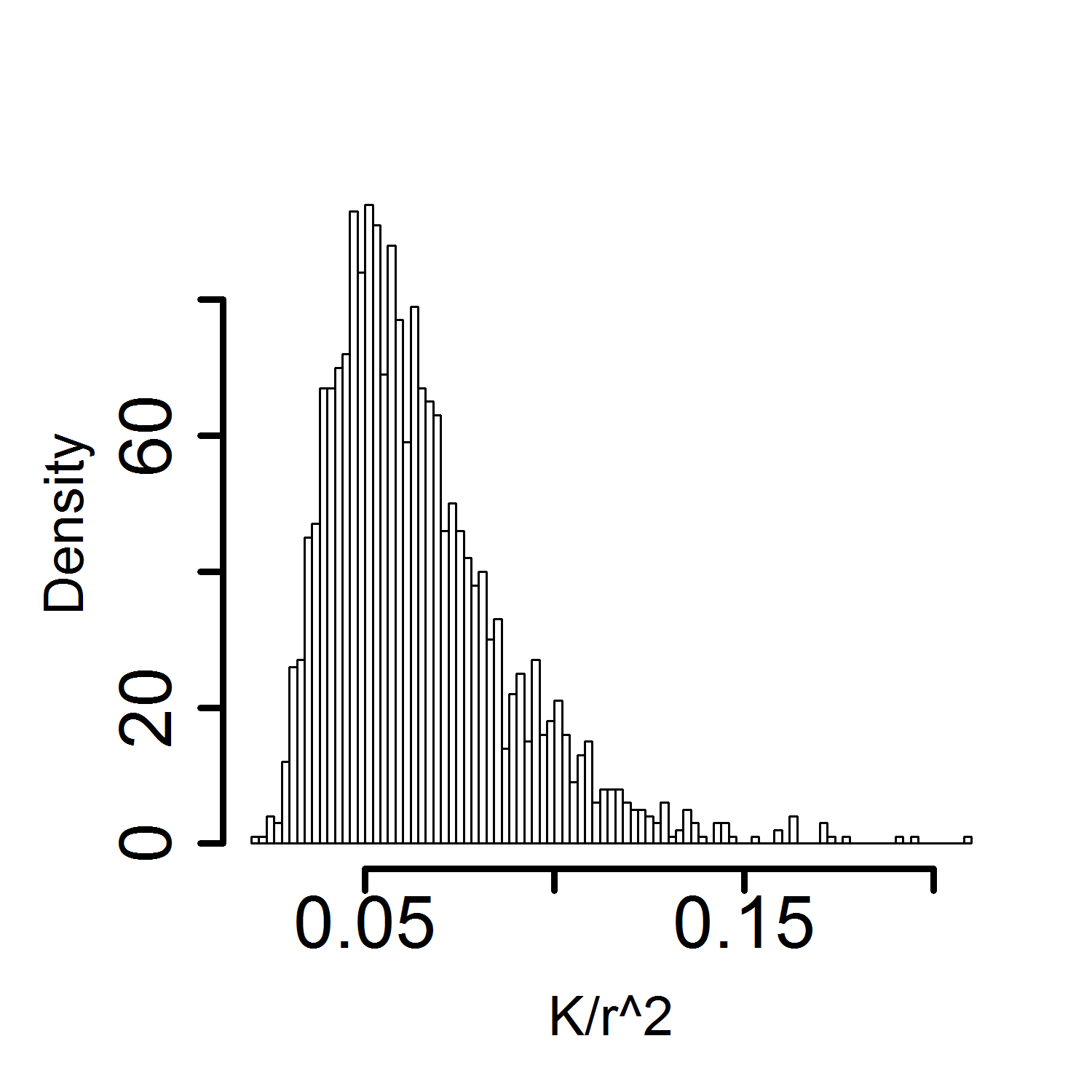}}\  &   1.78\ & 9.006\  \\
 \hline
   \end{tabular}}
\label{com}
\end{table}

\begin{figure}[h!]
\centerline{\includegraphics[scale=0.5]{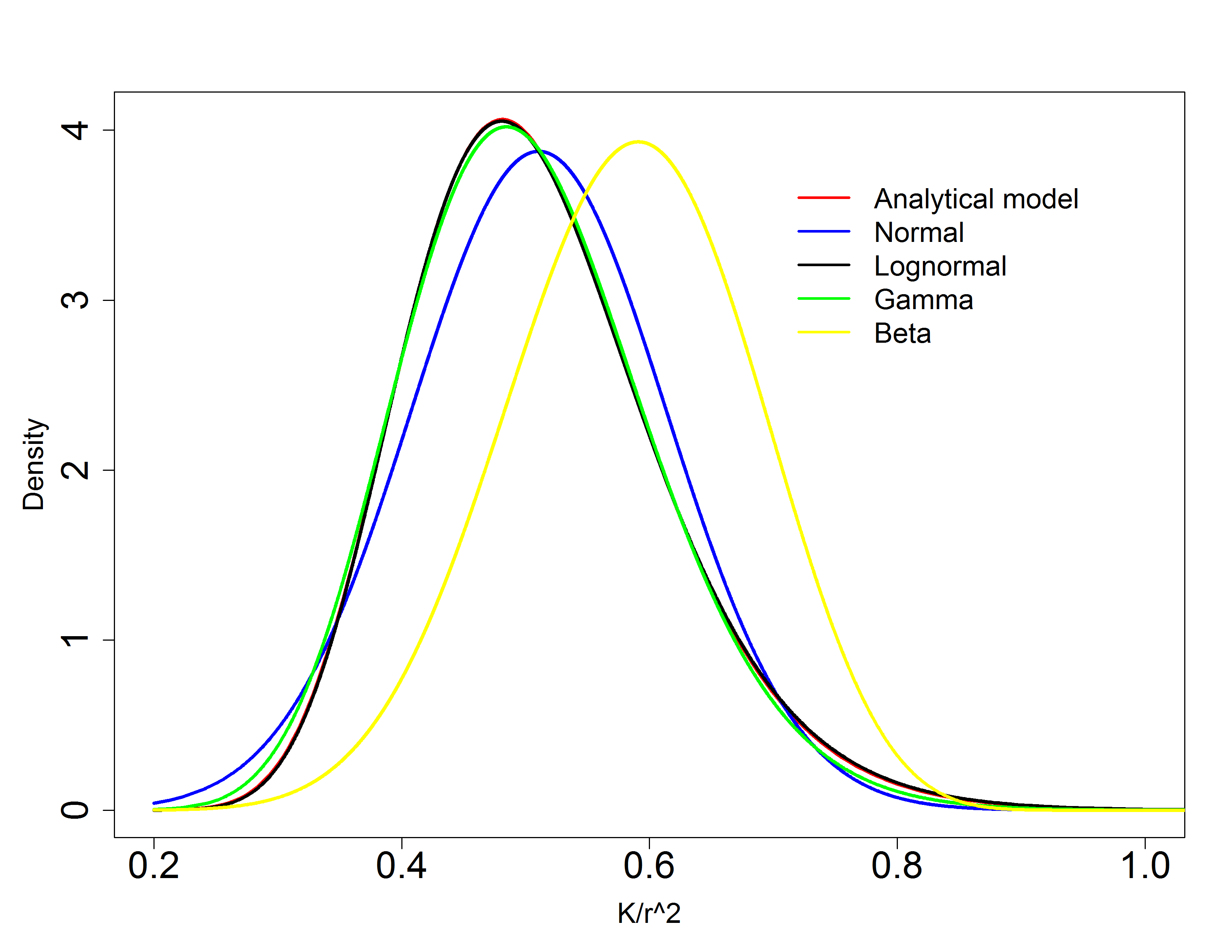}}
\caption{Comparing the computed PDF with common desity functions}
\label{FA}
\end{figure}

Furthermore, Figure~\ref{FA} illustrates how well the different distributions including normal, lognormal, gamma, and beta fit to the analytical probability function. It is clear that the  normal and beta distributions fail to represent the distribution of permeability values.
Figure~\ref{FA} also shows a lognormal distribution as well as a gamma distribution provide a good fit for permeability values. It has been derived that both lognormal and gamma distributions can be used effectively in analyzing a non-negative right-skewed data set \cite{cu}. As shown in Figure~\ref{FA}, a comparison between the lognormal and gamma distributions reveals that  their  PDFs have similar results. To find which of these distributions gives better fit to the data, we have considered two data transformation methods, one based on the normal approximation to the log of the data set, working on lognormal distributions and the other based on the cube root of the data, working on gamma distribution: If the data looks symmetric after log transformation, the lognormal distribution would work better to represent the variation of the permeability. If the data looks symmetric after the cube root transformation, the gamma distribution would work better to represent the variation of the permeability \cite{wh}.
\begin{figure}[h!]
\centerline{\includegraphics[scale=0.5]{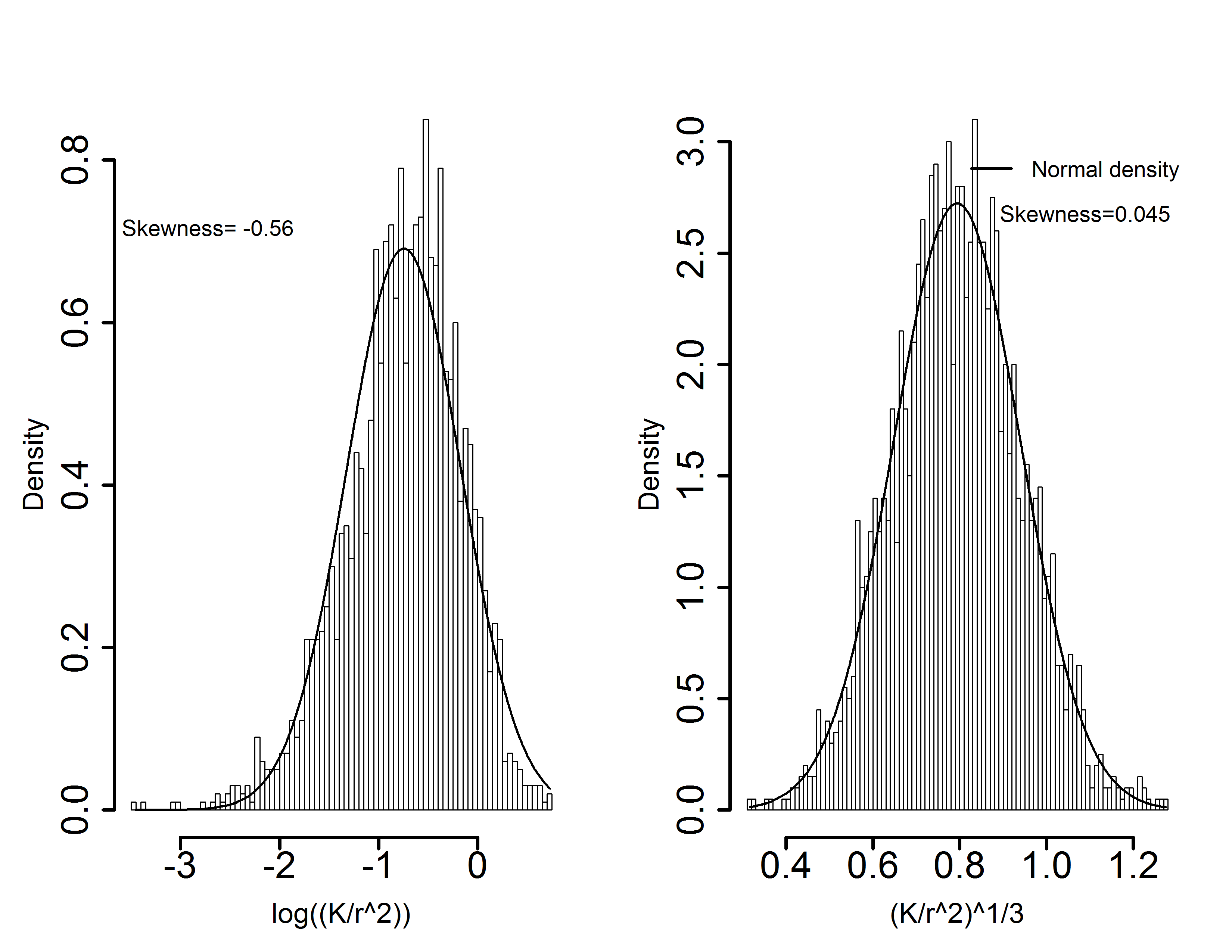}}
\caption{ Fitted normal densities for the two scenarios of transformation. The plot on the left side is based on logtransformation and the plot on the right side is based on the cube root transformation}
\label{gl}
\end{figure}

\begin{figure}[h!]
\centerline{\includegraphics[scale=0.5]{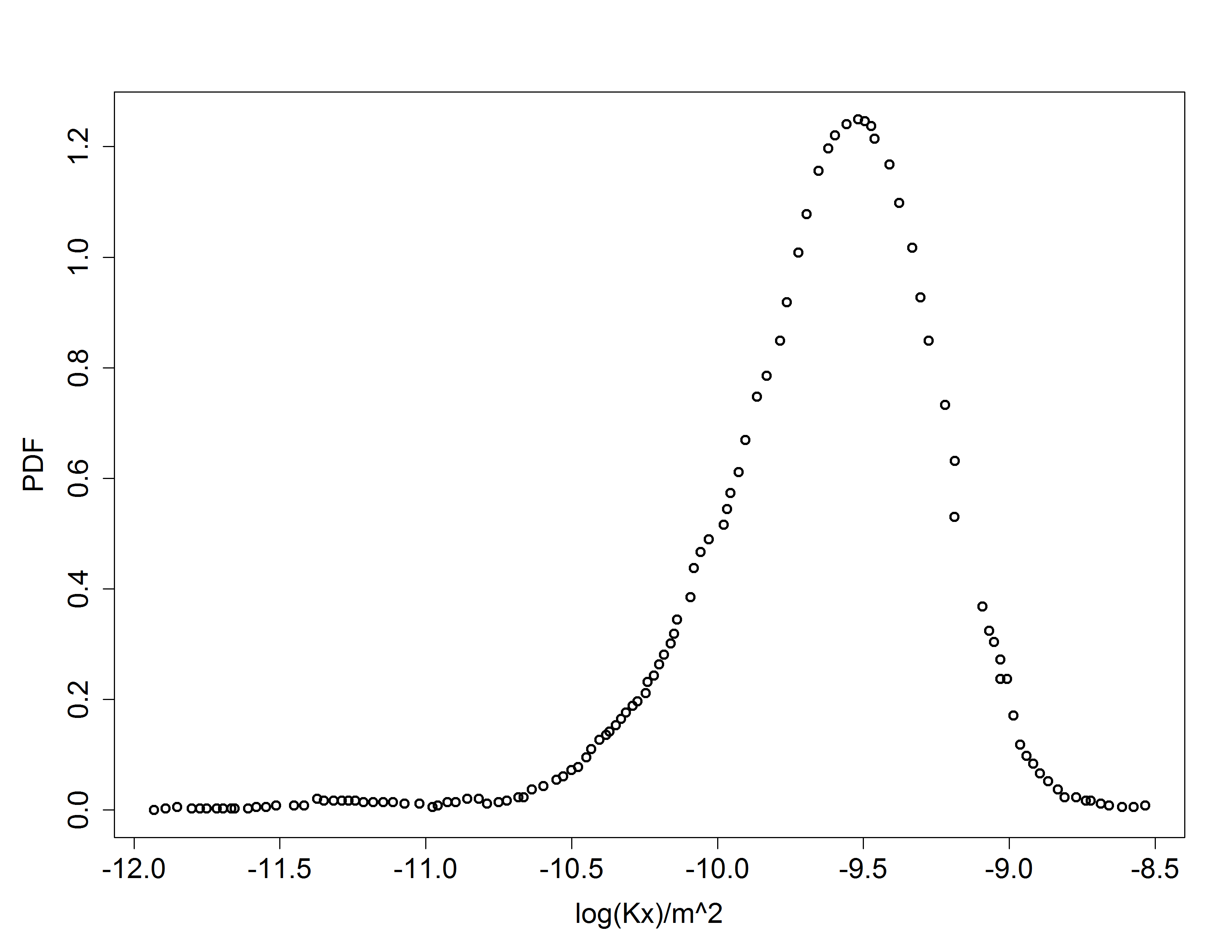}}
\caption{ PDF of permeability\cite{SMP}}
\label{zh}
\end{figure}

According to Figure~\ref{gl}, although the log transformation seems to fit well in the body of permeability values, the data shows to be left-skewed on log scale. The significant left-skewed on log scale also is observable on the work of  Zhang et al.\cite{SMP}(see Figure~\ref{zh}), who evaluated the permeability from local areal weight combined with the Kozeny-Carmen model, suggesting that the lognormal distribution can be used to describe the permeability distribution.


One final comment on which we would like to conclude this section regards entropy. In the literature it has been proven that under constraints, known mean and variance,  the normal distribution maximizes entropy. Using this principle that maximizes entropy is a selecting factor for a model we calculated the entropy values of the two fits.  Using straightforward computations we obtain the entropies for lognormal PDF -0.931 and gamma PFD -0.887, which coincids with the previous result claiming gamma distribution with more entropy determines the permeability data


.

\subsubsection{ Applicability of the gamma distribution}
In order to examine the applicability of gamma distributions to other empirical permeability equations, these equations are subjected to the Kolmogorov-Smirnov Statistics (KSS) test for normality in terms of normal, lognormal, gamma, weibull and beta distributions. If, in the KSS test P <0.05, there is significant probability of deviation from normality. The P value computations are listed in Table~\ref{gamma}. Gamma distribution shows the largest P-values among the given distributions for the different empirical permeability equations. Hence, based on the information in Table~\ref{gamma}, we conclude that the gamma distribution provides the best fit. Furthermore, it is shown in Table~\ref{gamma} the permeability COVs of all empirical equations range between 0.43 and 0.74, which is about 8 times larger than the fibre volume fraction COV of 0.086. This suggests that permeability is subjected to larger uncertainty than fibre volume fraction.

\begin{table}[!htbp]
\caption{Applicability of gamma distribution for different empirical equations}
\centering
\resizebox{\textwidth}{!}{\begin{tabular}{ |l|l|l|l|l|l|l|}
    \hline
   \multicolumn{1}{|c}{References} & \multicolumn{1}{|c|}{Permeability COV} & \multicolumn{5}{|c|}{KSS test}  \\ \cline{3-7}
     &  & Normal & Lognormal & Gamma & Weibull & Beta \\
    \hline
    Kozeny-Carmen (1937)\cite{CCP} & 0.475\  & 4.67e-13\ & 4.435e-5\ & 0.346\ & 2.97e-06\ & 1.017e-09\ \\
    \hline
   Gebart(Square) (1992)\cite{geba} & 0.554\ &  6.106e-15\ & 1.285e-8\ &0.05244\ & 0.0003844\ & 2.2e-16\ \\
    \hline
   Gebart(Hexagonal) (1992)\cite{geba} & 0.439\ &4.902e-11\ & 4.96e-6\ & 0.1171\ & 1.526e-05\ & 0.04888\ \\
    \hline
   Bruschke and Advani (1993)\cite{BRU}& 0.52\ & 4.6e-14\  &0.000114\ &  0.2883\ & 1.307e-6\ & 1.58e-8\ \\
 \hline
   Gutowski et al. (1987)\cite{GUT} & 0.739\ &  <2.2e-16\ &2.213e-9 \ &0.2948\ &0.0007611 \ & 0.04888\ \\
 \hline
   Happel (1959)\cite{HAP}& 0.44\  &2.112e-8\  &1.32e-5\ &0.1697\ &.0003369\ & 0.004764\ \\
 \hline
  Lee and Yang (1997)\cite{lee}& 0.53\  & 7.809e-12\  & 1.62e-7\ & 0.1209\ & 0.0031\ &  0.000438\ \\
 \hline
 Sharaoui and Kaviany (1992)\cite{sha}& 0.54\  &   <2.2e-16\  &  < 2.2e-16\ & 0.2336\ & 0.0018\ &  0.000238\ \\
 \hline
   \end{tabular}}
\label{gamma}
\end{table}

\section{Conclusions}
An adequate representation of microstructural variability of fibre arrangement in fibre-reinforced composites is of critical importance for the analysis of the flow in the fibrous media. The Distribution of fibre volume fraction was quantified by the measurement of areal weight density and areal thickness from optical images of tows in a $2\times2$ twill carbon-epoxy composite. Then, the PDF of the permeability was determined by a known PDF of $V_f$  and assuming a constant $k_c$ . To do so, we proposed a method to determine the probability density function of the permeability of porous media. We employed the Kozeny-Carmen equation and combined it with the change of variable technique. Our results suggest that
 $(1).$ The relationship between the local areal weight density and thickness is well approximated by a bivariate normal distribution.
 $(2).$ The distribution of local fibre volume fraction exhibits a bell-shaped curve and fit well to a normal distribution model.
 $(3).$ Assuming constant $k_c$, a gamma distribution could more accurately describe the variation in permeability data. \\
As conclusion, the understanding of the probability distribution of permeability is still taking further clarification but that the hypothesis of normality has been refuted.

\section*{Acknowledgement}
SYS thanks the CAPES PNPD-UFSCAR Foundation
for the financial support in the year 2014-5. SYS thanks the Federal University of Sao Carlos, Department of Statistics, for hospitality in 2014-5.

\newpage


\end{document}